\newif\ifOneCol

\OneColtrue

\ifOneCol

\documentclass[draftclsnofoot,twoside,onecolumn,letter,12pt]{IEEEtran}
\else
\documentclass[twocolumn,10pt]{IEEEtran}
\fi

\usepackage{cite}
\usepackage{graphicx}
\usepackage{amsmath}
\usepackage{amsfonts}
\usepackage{epstopdf}
\usepackage{xcolor}
\usepackage{amssymb,bm,upgreek}
\usepackage{algorithm}
\usepackage{algorithmic}
\usepackage{multirow}
\usepackage{transparent}
\usepackage{verbatim}
\usepackage{subfigure}
\usepackage{hyperref}
\hypersetup{
    colorlinks,
    citecolor=black,
    filecolor=black,
    linkcolor=black,
    urlcolor=black
}

\usepackage{import}
\usepackage{booktabs}

\DeclareMathOperator\erf{erf}
\DeclareMathOperator*{\E}{\mathbb{E}}

\newcommand{\prob}{\textnormal{Pr}}
\newcommand{\agg}{\textnormal{agg}}
\newcommand{\s}{\textnormal{s}}
\newcommand{\mole}{\textnormal{molecule}}

\newcommand{\m}{\textnormal{m}}
\newcommand{\self}{\textnormal{self}}

\newcommand{\im}{\textnormal{im}}
\newcommand{\var}{\textnormal{Var}}

\newcommand{\thmref}[1]{Theorem~\ref{#1}}
\newcommand{\proref}[1]{Proposition~\ref{#1}}

\newtheorem{remark}{Remark}
\newtheorem{theorem}{Theorem}
\newtheorem{corollary}{Corollary}
\newtheorem{proposition}{Proposition}
\newtheorem{lemma}{Lemma}
\newcommand{\Fr}[1]{{\color{black}{#1}}}

\begin{document}

\title{Characterization of Cooperators in Quorum Sensing with 2D Molecular Signal Analysis}

\author{Yuting Fang,
        Adam Noel,
        Andrew W. Eckford,
        Nan Yang,
        and
        Jing Guo
        \vspace{-10mm}

\thanks{This work was presented in part at IEEE Globecom 2019~\cite{GC2019}.}

}



\maketitle
\vspace{-4mm}
\begin{abstract}

In quorum sensing (QS), bacteria exchange molecular signals to work together. An analytically-tractable model is presented for characterizing QS signal propagation within a population of bacteria and the number of responsive cooperative bacteria (i.e., cooperators) in a two-dimensional (2D) environment. Unlike prior works with a deterministic topology and a simplified molecular propagation channel, this work considers continuous emission, diffusion, degradation, and reception among randomly-distributed bacteria. Using stochastic geometry, the 2D channel response and the corresponding probability of cooperation at a bacterium are derived. Based on this probability, new expressions are derived for the moment generating function and different orders of moments of the number of cooperators. The analytical results agree with the simulation results obtained by a particle-based method. In addition, the Poisson and Gaussian distributions are compared to approximate the distribution of the number of cooperators and the Poisson distribution provides the best overall approximation. The derived channel response can be generally applied to any molecular communication model where single or multiple transmitters continuously release molecules into a 2D environment. The derived statistics of the number of cooperators can be used to predict and control the QS process, e.g., predicting and decreasing the likelihood of biofilm formation.

\end{abstract}
\vspace{-4mm}
\begin{IEEEkeywords}
\vspace{-4mm}
Quorum sensing, molecular communication, 2D channel response, cooperative bacteria \vspace{-0mm}
\end{IEEEkeywords}

\IEEEpeerreviewmaketitle

\vspace{-2mm}
\section{Introduction}\label{sec:intro}
Quorum sensing (QS) is a ubiquitous approach for microbial communities to respond to a variety of situations in which monitoring the local population density is beneficial. When bacteria use QS, they assess the number of other bacteria they can interact with by releasing and recapturing a molecular signal in their environment.
This is due to the fact that a higher density of bacteria leads to more molecules that can be detected before they diffuse away or become degraded. If the number of molecules detected exceeds a threshold, then bacteria express target genes for a cooperative response. QS enables coordination within large groups of cells, potentially increasing the efficiency of processes that require a large population of cells working together.
Microscopic populations utilize QS to complete many collaborative activities, such as virulence, bioluminescene, biofilms, and the production of antibiotics. These tasks play a crucial role in bacterial infections, environmental remediation, and wastewater treatment \cite{7506074}.
Since the QS process is highly dependent on signaling molecules, an accurate characterization of release, diffusion, degradation, and reception of such molecules across the environment in which bacteria live is very important to understand and control QS, which can help us to prevent undesirable bacterial infections and lead to new environmental remediation methods \cite{0006655}.



There are growing research efforts to study the coordination of bacteria via QS. Among them, \cite{0006655,WEST2007,Lindsay2017WhenIP} investigated the cooperative behavior of bacteria using simulation or biological experiments and \cite{8278046,7181698,7935509,8422668,7397847,7742378} mathematically modeled bacterial behavior coordination.
We note that \cite{8278046,7181698,7935509,8422668,7397847,7742378} relied on abstract or simplifying models to represent the molecular diffusion channel (e.g., they did not consider the motion of individual signaling molecules based on Fick's laws) in order to focus on understanding how behavior evolves over time. \cite{6648629} considered a molecular diffusion channel between two clusters of bacteria, but did not consider the bacteria behavioral response. To the best of our knowledge, analyzing the statistics of the number of cooperative bacteria, taking into account the chemical reaction and diffusion of each molecule based on reaction-diffusion dynamics, is not available in the literature.




To control cooperative bacterial activities (e.g., biofilm formation and bioluminescene) for medical treatment and environmental monitoring, an analytically-tractable model needs to be developed since it predicts the cooperative behavior of bacteria, considering their noisy molecular signal propagation. We achieve this goal in this paper by leveraging the knowledge of QS, mass diffusion, stochastic geometry, and probability processes. Since our model accounts for the random motion of molecules based on reaction-diffusion equations, our model can be used to predict and control the impact of environment parameters, e.g., diffusion coefficient, reaction rate, and population density, on the concentration of molecules observed by bacteria and the statistics of the number of responsive cooperative bacteria.


In this paper, we consider a two-dimensional (2D) environment over which the bacteria are randomly spatially distributed according to a point process model. This is motivated by that bacteria may move in realistic environments and their locations may be not fixed. In the point process model, 
the locations of bacteria are changing between realizations \cite{Baccelli2010,5226957}.  
This means that the number of cooperative bacteria can change from one realization to the next. 
As a result, we are interested in the average result or distribution of the number of cooperative bacteria over a large number of realizations. 
We consider a 2D environment since a 2D environment facilitates future experimental validation of our current theoretical work. Biological experiments, especially with bacteria, are usually conducted in a 2D environment, e.g., bacteria residing on a petri dish (i.e., a thin plate for cell-culture) and the formation of biofilms \cite{biofilm}. While considering the topological randomness of bacteria, our model captures the basic features of QS by adopting the assumptions as follows: We assume that each bacterium acts as both a point transmitter (TX) and a circular receiver (RX), which captures the features of emission and reception of QS molecules. Since bacteria emit molecules sporadically in reality, we assume that each bacterium continuously\footnote{Note that continuous emission does not mean there is no time interval between two successive emissions of molecules.} 
emits molecules at random times. 

We emphasize that developing the analytical model in this paper is theoretically challenging since we need to address the \emph{random} received signal at bacteria in \emph{random} locations due to \emph{randomness} in the motion and degradation of molecules, and \emph{randomness} in the locations of many TXs. Despite these challenges, we make the following theoretical contributions:
\begin{enumerate}
\item We analytically derive the channel response (i.e., the expected number of molecules observed) at a RX due to continuous emission or an impulse emission of molecules at one point TX. Based on this, we then derive the channel response at a RX due to continuous emission of molecules from point TXs randomly distributed over a circle in a 2D environment. 
\item Using the results in 1), we first derive the exact expression for the expected probability of cooperation at the bacterium at a fixed location, due to the emission of molecules from randomly-distributed bacteria, by using the Laplace transform of the random aggregate of received molecules. We then derive an approximate expression for such a probability, which is easier to compute than the exact expression, yet from our numerical results has good accuracy when the population density is sufficiently high.
\item Based on the results in 2), we derive approximate expressions for the moment generating function (MGF) and cumulant generating function (CGF) of the number of cooperative bacteria (i.e., cooperators). Using the MGF and the CGF, we derive approximate expressions for the $n$th moment and cumulant of the number of cooperators. We study the convergence of the number of cooperators to a Gaussian distribution via the higher order statistics. We use the Poisson and Gaussian distributions with the derived statistics to analytically fit the probability mass function (PMF) and cumulative distribution function (CDF) of the number of cooperators. We show that the Poisson distribution provides the best overall approximation, based on our numerical results. In addition, we derive the expected number of pairs of two nearest bacteria both cooperating, which can be used to predict clusters of cooperators in a QS system.
\end{enumerate}
We validate the accuracy of our analytical results via a particle-based simulation method where we track the random walk of each signaling molecule over time. In contrast to our preliminary work in \cite{GC2019}, which only derives a portion of the results in 1) and the expected number of cooperators, this paper conducts a more comprehensive analysis of the 2D channel response, derives an exact expression for the expected probability of cooperation, and studies the distribution of the number of cooperators by its MGF and different order statistics.

\Fr{Our derived statistical moments can help predict and control the QS process. For example, biofilm formation via QS is a mechanism for bacteria to resist antibiotics. However, a biofilm could be prevented from forming if the density of cooperators is too small. Our expressions reveal the impact of environmental factors (e.g., degradation and diffusion rates) on the likelihood of a given number of bacteria choosing to cooperate. Based on our expressions, we could infer how to decrease the likelihood of successful biofilm formation by adjusting environmental parameters. This could help optimize the performance of antibiotic treatment.}

\Fr{While contributing to QS, our results could also be applied to other molecular communication (MC) systems. More specifically, the 2D channel response could be used to determine the expected molecular signal observed at an RX when a TX (or TXs) impulsively or continuously releases molecules into a 2D environment in an MC system. The cooperating probability and statistical moments of the number of cooperators can be applied to study the group behavior of an MC system that uses consensus algorithms and broadcast channel models. For example, a cluster of nanomachines in a nanonetwork could secrete and sense molecules to achieve global network synchronization as proposed in \cite{Abadal2011}. Our derived cooperating probability and the number of cooperators could be used to determine the probability of a nanomachine being synchronized and the percentage of synchronized nanomachines in the nanonetwork.
} 



We use the following notations: $|\vec{x}|$ denotes Euclidean norm of a vector $\vec{x}$. $\overline{N}$ denotes the mean of a random variable (RV) $N$ and ${\E}_{\Phi}\{\cdot\}$ denotes the expectation over a spatial random point process $\Phi$. $K_n(\cdot)$ denotes modified $n$th order Bessel function of the second kind. $I_n(\cdot)$ denotes the modified $n$th order Bessel function of the first kind \cite{gradshteyn2007}. $\Gamma(a,z)$ denotes the incomplete gamma function.

\section{System Model}\label{sec:system model}

\begin{figure}[!t]
\centering
\includegraphics[width=0.5\columnwidth]{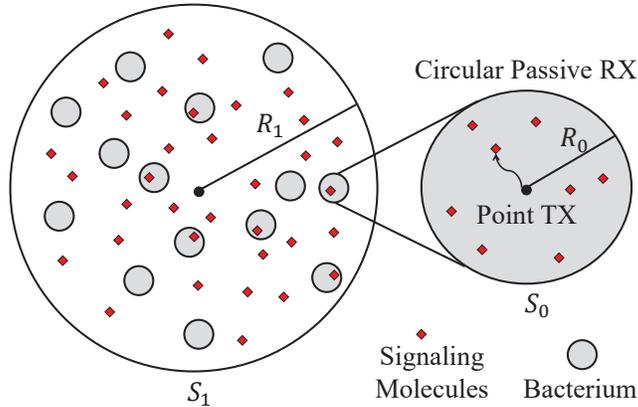}
\vspace{0mm}
\caption{A population of bacteria randomly distributed over a circle $S_1$ according to a 2D spatial point process, where each bacterium acts as a point TX and as a circular passive RX $S_0$. The molecules diffuse into and out of the bacteria.}
\label{fig:model}
\vspace{-2mm}
\end{figure}

%

We consider an \emph{unbounded} 2D environment. Unlike a deterministic topology model, we consider a point process model to represent topological randomness of bacteria. There are several types of point processes and we consider the Poisson point process (PPP) in this paper\cite{Haenggi2009} due to its tractable properties and well-known theorems. A population of bacteria is spatially distributed over a \emph{bounded} circle $S_1$ with radius $R_1$ centered at $(0,0)$ according to a 2D PPP with constant density $\lambda$, as shown in Fig.~\ref{fig:model}. 
\Fr{We denote $\vec{x_i}$ as the location of the center of the bacterium $i$ where the bacterium $i$ is an arbitrary bacterium in the bacterial population.} We denote $\Phi\left(\lambda\right)$ as the set of random bacteria locations. We consider bacteria behavior analogous to QS, i.e., 1) emit signaling molecules; 2) detect the concentration of signaling molecules; and 3) decide to cooperate if the concentration exceeds a threshold. In the following, we detail the emission, propagation, and reception of signaling molecules, and the decision-making by the bacteria.

\begin{figure*}[!tbp]
		\centering
		\begin{minipage}[t]{0.49\textwidth}\hspace*{-5 mm}
			\centering
			\resizebox{1.05\linewidth}{!}{
				\includegraphics[scale=0.55]{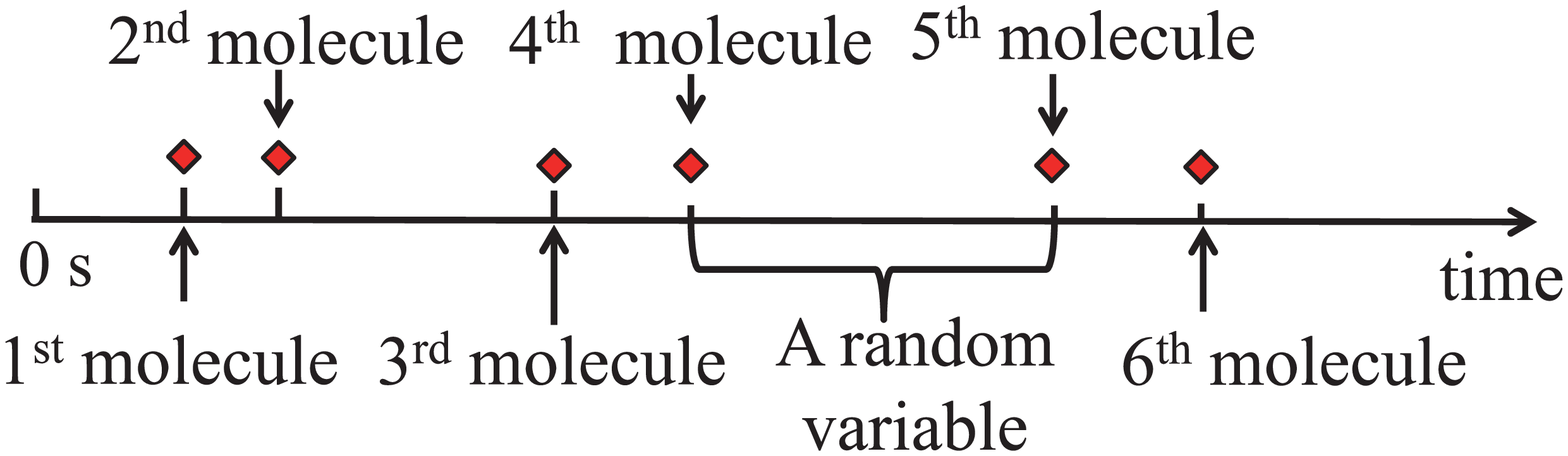}}\vspace*{-4 mm}
			\caption{An example of release times due to continuous emission of molecules at a bacterium according to a random process.}
			\label{fig:emit}
		\end{minipage}
		\hfill
		\vspace*{-1 mm}
		\begin{minipage}[t]{0.49\textwidth}
			\centering
			\resizebox{1.05\linewidth}{!}{
				\includegraphics[scale=0.55]{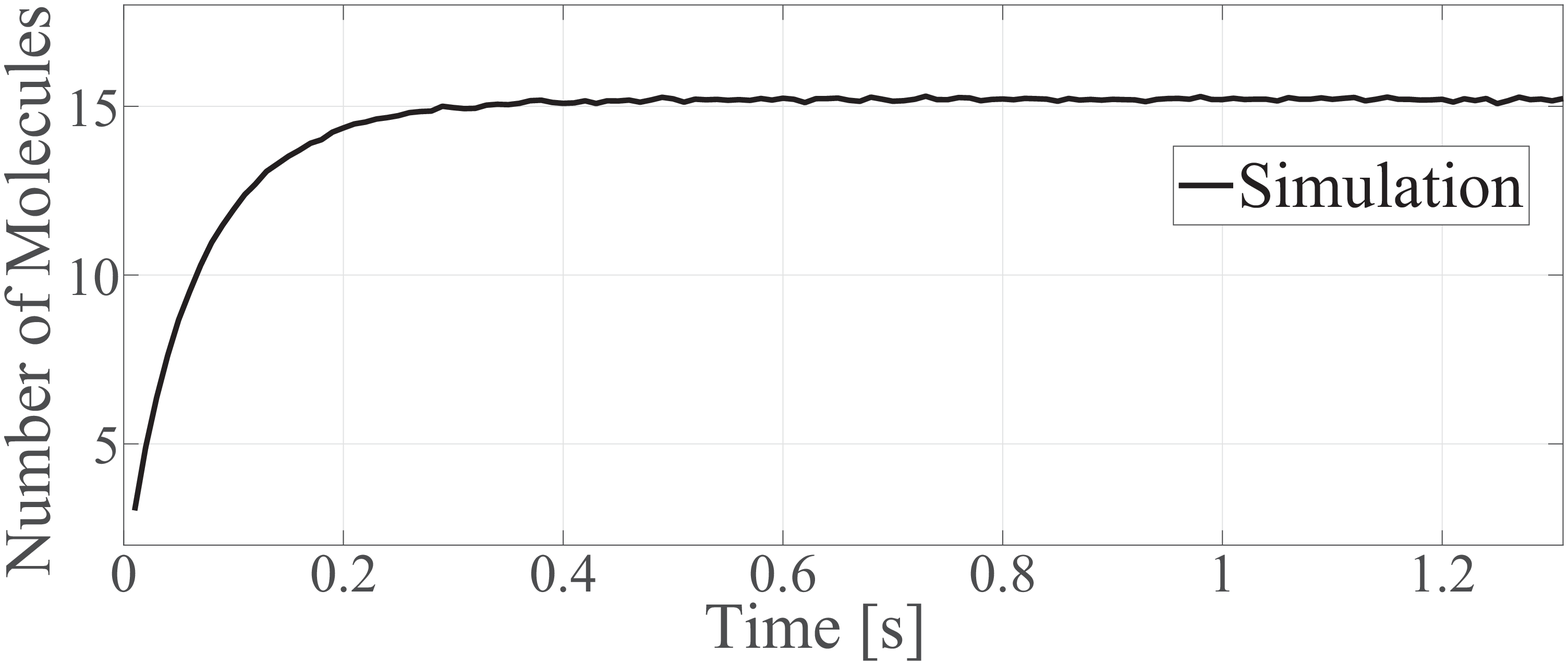}}\vspace*{-4 mm}
			\caption{The average number of molecules observed at time $t$, $\overline{N}_{\agg}^{\dag}\left(\vec{x_i},t|\lambda\right)$, versus time $t$. $R_1 = 20\,{\mu}\m$, $\lambda=7.9\times10^{-2}/{\mu}\m^2$,  $\vec{x_i}=(10\,{\mu}\m,10\,{\mu}\m)$; see other simulation details in Sec. \ref{sec:Numerical}.}%
			\label{radiusAndtime}
		\end{minipage}
		\vspace*{-1 mm}
		
	\end{figure*}

\Fr{\textbf{Emission}: We consider that bacterium $i$ continuously releases molecules at different times, as shown in Fig. \ref{fig:emit}. The release time instants constitute points in a one-dimensional (1D) PPP and one molecule is released at each release time instant. We assume that the continuous emission of each bacterium follows an independent PPP with a rate $q$. The rate $q$ is the average number of release time events within a unit time (i.e., 1 second). The rate $q$ is also equal to the number of molecules being released per second since one molecule is released at each release time event. We note that \cite{6949026} and \cite{8640823} also model molecules being emitted continuously at random times according to a 1D PPP.}



\textbf{Propagation}: All $A$ molecules diffuse independently with a constant diffusion coefficient $D$ and they can degrade into a form that cannot be detected by the bacteria, i.e., $A\overset{k}\rightarrow\emptyset$, where $k$ is the reaction rate constant in $\s^{-1}$. If $k=0$, this degradation is negligible. Since we consider a single type of molecule, we only mention ``the molecules'', i.e., omitting ``$A$'', in the remainder of this paper.

\textbf{Reception}: \Fr{We model the bacterium $i$ as a circular passive RX with radius $R_0$ and area $S_0$ centered at $\vec{x_i}$. The bacterium $i$ samples the number of molecules within $S_0$ at only \textbf{only one time instant $t$}.} Bacteria perfectly count molecules if they are within $S_0$. Since the molecules released from all bacteria may be observed by the bacterium $i$, the number of molecules observed at the bacterium $i$ at time $t$, ${N}_{\agg}^{\dag}\left(\vec{x_i},t|\lambda\right)$, is given by $N_{\agg}^{\dag}\left(\vec{x_i},t|\lambda\right)=\sum_{\vec{x_j}\in\Phi\left(\lambda\right)}N\left(\vec{x_i},t|\vec{x_j}\right)$, where $N\left(\vec{x_i},t|\vec{x_j}\right)$ is the number of molecules observed at time at the bacterium $i$ due to the bacterium $j$. The means of $N_{\agg}^{\dag}\left(\vec{x_i},t|\lambda\right)$ and $N\left(\vec{x_i},t|\vec{x_j}\right)$ are denoted by $\overline{N}_{\agg}^{\dag}\left(\vec{x_i},t|\lambda\right)$ and $\overline{N}\left(\vec{x_i},t|\vec{x_j}\right)$, respectively. We assume that the \emph{expected} number of molecules observed at the bacterium $i$ is constant after some time when degradation occurs. To demonstrate the suitability of this assumption, see Fig. \ref{radiusAndtime} (and an analytical proof in Remark \ref{re:asymptotic}). In Fig. \ref{radiusAndtime}, $\overline{N}_{\agg}^{\dag}\left(\vec{x_i},t|\lambda\right)$ is independent of $t$ after time $t\approx0.5\,\s$. We denote time $t^{\star}_i$ as the time after which $\overline{N}_{\agg}^{\dag}\left(\vec{x_i},t|\lambda\right)$ is approximately constant, i.e.,
\begin{align}\label{asymp}
\overline{N}_{\agg}^{\dag}\!\left(\vec{x_i},t|\lambda\right)\!|_{t>t^{\star}_i}\!\approx\!\lim_{t\rightarrow\infty}\!\overline{N}_{\agg}^{\dag}\!\left(\vec{x_i},t|\lambda\right)\!=\!\overline{N}_{\agg}^{\dag}\!\left(\vec{x_i},\infty|\lambda\right).
\end{align}
We also refer to any observation after $t^{\star}_i$, i.e., $\overline{N}_{\agg}^{\dag}\!\left(\vec{x_i},\infty|\lambda\right)$, as an asymptotic observation. \Fr{Notably, similar steady observations at the asymptotic stage have also been shown in \cite{6949026} and \cite{8640823}.}


\textbf{Decision-Making}: We assume that the bacterium $i$ uses its asymptotic observation to make a decision, when the expected number of observed molecules becomes stable. This assumption is reasonable since $t^{\star}_i$ is very small, e.g., $t=0.5\,\s$ in Fig. \ref{radiusAndtime}, and bacteria can reach the steady state very quickly, especially relative to the timescale of gene regulation to coordinate behavior\footnote{Based on \cite{Danino2010,Trovato2014,Surette7046}, the cooperation of bacteria is observed after the signaling molecules diffuse for at least tens of minutes.}. Also, bacteria can wait until there are enough molecules to trigger behavior change. Therefore, bacteria do not need to explicitly know whether the steady state has been reached and precise synchronization over the population for emission and detection is not needed. 
Inspired by QS, we consider a threshold-based strategy at bacteria to decide cooperation. We note that the threshold-based strategy is commonly adopted in molecular communication (MC) literature, e.g., \cite{Tepekule2015,Lin2015}. We consider that bacteria compare $N_{\agg}^{\dag}\left(\vec{x_i},\infty|\lambda\right)$ with a threshold $\eta$. If $N_{\agg}^{\dag}\left(\vec{x_i},\infty|\lambda\right)\geq\eta$, then the bacterium $i$ decides to be a cooperator, otherwise the bacterium $i$ is noncooperative.  

In the following, we acknowledge the major simplifying assumptions to clarify the applicability of our work and identify areas for future study. 
These assumptions are as follows:
\begin{enumerate}
\item  \Fr{We assume that the bacteria do not move over time after being randomly placed in each realization of the 2D PPP. This assumption is reasonable for three reasons: i) There are some non-motile bacteria, e.g., coliform and streptococci; ii) When bacteria swim very slowly, the mobile case can be well approximated by the non-mobile case; and iii) In fact, bacteria often keep stationary when cooperating, e.g., when forming a stable biofilm.}

\item We consider an ideal transceiver model for bacteria. We simplify bacteria as point sources emitting molecules isotropically into the environment. Considering imperfect TXs is left for future work. We assume that bacteria are passive observers that do not interact with signaling molecules. This is because the observations at multiple bacteria are \emph{correlated} for reactive RXs, which makes analysis much more cumbersome.
\item We assume that the average molecule emission rate is constant. We acknowledge that in a real QS process, bacteria may increase their emission rate when they change from being selfish to being cooperative. This assumption is appropriate for scenarios where bacteria transition from being selfish to ramping up molecule production before the emission rate is updated.

\item Each bacterium makes one decision based on one sample of the observed signal. We acknowledge that bacteria usually make decisions to cooperate multiple times in their life. Modeling evolutionary or repeat behavior coordination over time with noisy signal propagation is interesting for future work, e.g., as we propose in \cite{Game_tutorial}.
\end{enumerate}

\section{2D Channel Response}\label{sec:2D}

In this section, we aim to derive the channel response, i.e., the expected number of molecules observed at a RX, due to continuous emission of molecules from randomly-distributed TXs. To this end, we first derive the channel response due to a point TX as an intermediate step. We assume that the RX is a circular passive observer $S_0$ centered at $\vec{b}$ with radius $R_0$ throughout this section, unless specified otherwise. These analyses lay the foundations for our derivations of the observations at bacteria and expected density of cooperators in Sec. IV. 

\subsection{One Point TX}

\Fr{In this subsection, we only consider a single TX-RX pair and the point TX is at the point (0,0) for the convenience of analysis. The scenario considered here for this preliminary analysis is distinct from our general system model where multiple TXs exist.} We also include the special case when the TX is at the center of the RX, since each bacterium receives the molecules released not only from other bacteria but also from itself.

Based on  \cite[eq.~(9)]{6949026}, the asymptotic channel response due to \emph{continuous} emission is obtained by multiplying the impulse channel response by the emission rate $q$ and then integrating over all time to infinity. Using this method, the asymptotic channel response is given in the following proposition:
\begin{proposition}\label{propo:impulse-to-cont}
The asymptotic channel response $\overline{N}\left(\vec{b},\infty\right)$ at $\vec{b}$, due to continuous emission with rate $q$ from the point $(0,0)$ since time $t=0$, is given by
\begin{align}\label{cont-impulse}
\overline{N}\left(\vec{b},\infty\right) = &\;q\int_{\tau=0}^{\infty}\overline{N}_{\im}\left(\vec{b},\tau\right)d\tau,
\end{align}
where $\overline{N}_{\im}\left(\vec{b},\tau\right)$ is the channel response at $\vec{b}$ at time $\tau$ due to an impulse emission of one molecule at time $t=0$ from the point $(0,0)$.
\end{proposition}
\begin{IEEEproof}
See Appendix~\ref{Proof of impulse to conti}.
\end{IEEEproof}

Based on Proposition \ref{propo:impulse-to-cont}, we first derive $\overline{N}_{\im}\left(\vec{b},\tau\right)$ for any $\vec{b}$ and $\vec{b}=0$ to evaluate $\overline{N}\left(\vec{b},\infty\right)$. We note that the results of $\overline{N}_{\im}\left(\vec{b},\tau\right)$ also can be used in any contexts where a TX emits molecules impulsively. The impulsive emission is commonly considered in the MC literature, which assumed that the TX is a nanomachine having the ability to control the timing of its molecule releases.

\subsubsection{Impulse Emission}
We first solve $\overline{N}_{\im}\left(\vec{b},\tau\right)$ for any $\vec{b}$ and then solve the special case when the TX is at the center of the circular RX $S_0$, i.e., $|\vec{b}|=0$. We denote $\overline{N}_{\im,\self}\left(\tau\right)$ as the channel response at $\vec{b}=(0,0)$ at time $\tau$ due to an impulse emission of one molecule at time $t=0$ from the point $(0,0)$, i.e., $\overline{N}_{\im,\self}\left(\tau\right) = \lim_{\vec{b}\to\mathbf{0}}\overline{N}_{\im}\left(\vec{b},\tau\right)$. We solve $\overline{N}_{\im}\left(\vec{b},\tau\right)$ for any $\vec{b}$ and $\overline{N}_{\im,\self}\left(\tau\right)$
in the following theorems.
\begin{theorem}[Impulse Emission for Any $\vec{b}$]\label{Theorem:impulse,any}
The channel response $\overline{N}_{\im}\left(\vec{b},\tau\right)$ for a circular passive observer $S_0$ centered at any $\vec{b}$ with radius $R_0$ is given by
\begin{align}\label{point-circle, impulse2}
\overline{N}_{\im}\left(\vec{b},\tau\right)
 = &\;\sum_{i=1}^{4}\bigg\{{\alpha_i}\exp\bigg(-\frac{{R_0}^2+{|\vec{b}|}^2}{4D\tau}-k\tau\bigg)\bigg[\exp\bigg(\frac{{R_0}^2}{4D\tau}\bigg)-\exp\bigg(\frac{{R_0}|\vec{b}|\beta_i}{2D\tau}\bigg)\bigg]\nonumber\\
&+\frac{\alpha_i |\vec{b}|\beta_i\sqrt{D \pi}}{2D\sqrt{\tau}}\exp\!\bigg(\!-\!\frac{{|\vec{b}|}^2(1\!-{\beta_i}^2)}{4D\tau}\!-k\tau\!\bigg)\! \bigg[\!\erf\bigg(\frac{|\vec{b}|\beta_i }{2\sqrt{D\tau}}\bigg)+\erf\bigg(\frac{R_0-|\vec{b}|\beta_i }{2\sqrt{D\tau}}\bigg)\!\bigg]\bigg\},
\end{align}
where the exact values of $\alpha_i$ and $\beta_i$ for different ranges of $z$ are given in \cite{Salahat2013}. Due to the limitation of space, we do not present these exact values here.
\end{theorem}
\begin{theorem}[Impulse Emission for $|\vec{b}|=0$]\label{Theorem:impulse,0}
The channel response due to an impulse emission from itself is given by
\begin{align}\label{point-circle,self2}
\overline{N}_{\im, \self}\left(\tau\right) =&\;\exp(-k\tau)\left(1-\exp\left(\frac{-R_0^2}{4D\tau}\right)\right).
\end{align}
\end{theorem}
\begin{IEEEproof}
The proofs of Theorem \ref{Theorem:impulse,any} and Theorem \ref{Theorem:impulse,0} are given in Appendix~\ref{Chap4:Proof of impulse}.
\end{IEEEproof}

\subsubsection{Continuous Emission}
We then evaluate the asymptotic channel response due to continuous emission for any~$\vec{b}$ and $\vec{b}=0$ in the following theorems. 
\begin{theorem}[Continuous Emission for Any $\vec{b}$]\label{Theorem:continuous,any}
The asymptotic channel response $\overline{N}\left(\vec{b},\infty\right)$ for the circular passive RX $S_0$ centered at any $\vec{b}$, due to continuous emission with rate $q$ from the point $(0,0)$ since time $t=0$, using uniform concentration assumption (UCA) \cite{Noel2013}, is given by
\begin{align}\label{point1}
\overline{N}\left(\vec{b},\infty\right)
\approx \frac{q R_0^2}{2D}K_0\left(|\vec{b}|\sqrt{\frac{k}{D}}\right).
\end{align}
\end{theorem}
\begin{theorem}[Continuous Emission for $|\vec{b}|=0$]\label{Theorem:continuous,0}
The asymptotic channel response at the circular RX $S_0$, due to continuous emission with rate $q$ from the center of this RX since time $t=0$, $\overline{N}_{\self}\left(\vec{b},\infty\right)$, is given by
\begin{align}\label{point-circle,self}
\overline{N}_{\self}\left(\infty\right)
= \frac{q}{k}\bigg(1-{\frac{\sqrt{k}R_0}{\sqrt{D}}K_1\bigg(\sqrt{\frac{k}{D}}R_0\bigg)}\bigg)
\end{align}
\end{theorem}
\begin{IEEEproof}
The proof of Theorem \ref{Theorem:continuous,any} and Theorem \ref{Theorem:continuous,0} are given in Appendix~\ref{Proof:Cont-any}.
\end{IEEEproof}

This UCA is accurate if $|\vec{b}|$ is relatively large and thus it is inaccurate when $|\vec{b}|=0$. The accuracy of the UCA applied in \eqref{point1} will be verified in Sec. \ref{sec:Numerical}.
\begin{remark}\label{re:asymptotic}
We have analytically found that $\overline{N}_{\agg}^{\dag}\left(\vec{x_i},t|\lambda\right)$ converges as time $t\rightarrow\infty$ when $k\neq0$, since $\overline{N}_{\agg}^{\dag}\left(\vec{x_i},t|\lambda\right)=\sum_{\vec{x_j}\in\Phi\left(\lambda\right)}\overline{N}\left(\vec{x_i},t|\vec{x_j}\right)$ and, from \eqref{point1}, $\overline{N}\left(\vec{x_i},\infty|\vec{x_j}\right)$ is a constant when $k\neq0$. This analytically proves that our assumption adopted for \textbf{Reception} in Sec. \ref{sec:system model} is valid, i.e., $\overline{N}_{\agg}^{\dag}\left(\vec{x_i},t|\lambda\right)$ does not vary with time $t$ after some time when $k\neq0$.
\end{remark}
\begin{remark}\label{re:asymptotic1}
We note that $\overline{N}\left(\vec{b},\infty\right)\to\infty$ when $k=0$ since $K_0(0)\to\infty$. Thus, $\overline{N}_{\agg}^{\dag}\left(\vec{x_i},t|\lambda\right)=\sum_{\vec{x_j}\in\Phi\left(\lambda\right)}\overline{N}\left(\vec{x_i},t|\vec{x_j}\right)$ does not converge as time $t\rightarrow\infty$ when $k=0$, which will be verified in Sec. \ref{sec:Numerical}. Since $\overline{N}\left(\vec{b},\infty\right)$ when $k=0$ does not converge, we evaluate the time-varying channel response $\overline{N}\left(\vec{b},t\right)$ at time $t$ with no molecule degradation, i.e., $k=0$, which is given by
\begin{align}\label{point,no deg}
\overline{N}\left(\vec{b},t\right)\Big{|}_{k=0} \approx \pi R_0^2\int_{\tau=0}^{t}\frac{q}{(4\pi D \tau)}\exp\left(-\frac{|\vec{b}|^{2}}{4D\tau}\right)d\tau
\approx\frac{\Gamma\left(0,\frac{|\vec{b}|^2}{4Dt}\right)qR_0^2}{4D}.
\end{align}
\end{remark}


\subsection{Randomly-Distributed TXs}\label{subsec:2D,TXs}
In this subsection, we consider that \emph{many} point TXs are randomly distributed over a circle $S_1$ according to a point process with a constant density $\lambda$, as shown in Fig. \ref{fig:model}. The circle $S_1$ is centered at $(0,0)$ with radius $R_1$. We represent $\vec{a}$ as the location of an arbitrary point TX $a$ and $\Phi\left(\lambda\right)$ as the set of TXs' random locations. We denote the asymptotic channel response at the circular RX $S_0$ centered at $\vec{b}$ with radius $R_0$, due to continuous emission with rate $q$ since time $t=0$ from TX $a$, by $\overline{N}\left(\vec{b},\infty|\vec{a}\right)$, and the corresponding aggregate channel response at the RX due to all randomly-distributed TXs on circle $S_1$ with density $\lambda$ by\footnote{In this paper, $N_{\agg}$ and $N_{\agg}^{\dag}$ denote the observation at a RX and at a bacterium, respectively. We use the superscript $\dag$ to differentiate whether the observation includes the molecules released from itself.} $\overline{N}_{\agg}\left(\vec{b},\infty|\lambda\right)=\sum_{\vec{a}\in\Phi\left(\lambda\right)}\overline{N}\left(\vec{b},\infty|\vec{a}\right)$. We denote ${\E}_{\Phi}\left\{\overline{N}_{\agg}\left(\vec{b},\infty|\lambda\right)\right\}$ as the expected $\overline{N}_{\agg}\left(\vec{b},\infty|\lambda\right)$ over the point process $\Phi\left(\lambda\right)$. For compactness, we remove $\infty$ in all notation in the remainder of this paper since we assume that bacteria use asymptotic observations of the continuous emission by TXs to make decisions. We next derive ${\E}_{\Phi}\left\{\overline{N}_{\agg}\left(\vec{b}|\lambda\right)\right\}$ and simplify it in different special cases in order to ease the computational complexity.

\begin{theorem}\label{Theorem:aggregate_mole}
The expected aggregate channel response at the RX due to all randomly distributed TXs on circle $S_1$ with density $\lambda$ over the point process $\Phi\left(\lambda\right)$ is given by
\begin{align}\label{circle-cirle, agg}
{\E}_{\Phi}\left\{\overline{N}_{\agg}\left(\vec{b}|\lambda\right)\right\}
=&\;\int_{|\vec{r}|=0}^{R_1}\int_{\varphi=0}^{2\pi}\overline{N}\left(\vec{b}|\vec{r}\right)\lambda |\vec{r}|\,d\varphi\,d|\vec{r}|\nonumber\\
= &\;\lambda \int_{|\vec{r}|=0}^{R_1}\int_{\varphi=0}^{2\pi}\int_{|\vec{r_0}|=0}^{R_0}\int_{\theta=0}^{2\pi}K_0\left(\sqrt{\frac{k}{D}} \Upsilon(\vec{b})\right)\frac{q}{2D\pi}|\vec{r_0}||\vec{r}|\,d\theta\,d|\vec{r_0}| \,d\varphi\,d|\vec{r}|.
\end{align}
where $\Upsilon(\vec{b})$ is given in
$\Upsilon(\vec{b}) = \sqrt{\Omega(\vec{b})+|\vec{r_0}|^2+2\sqrt{\Omega(\vec{b})}|\vec{r_0}|\cos\theta}$
and $\Omega(\vec{b})={|\vec{b}|}^2+|\vec{r}|^2+2{|\vec{b}|}|\vec{r}|\cos\varphi$.
\end{theorem}
\begin{IEEEproof}
See Appendix~\ref{Chap4:aggregate_mole}.
\end{IEEEproof}

Although we consider a point TX, a circular TX is also of interest since a realistic TX, e.g., a cell, could be approximately spherical and molecules can be generated in different sections throughout a cell. We discuss the channel response due to a circular TX in the following remark:
\begin{remark}\label{re:circluarTX}
It can be shown that the asymptotic channel response at a circular RX with radius $R_0$ centered at $\vec{b}$, due to continuous emission with rate $q$ from a \emph{circular} TX centered at $(0,0)$ with radius $R_1$ since time $t=0$, can be obtained by removing density $\lambda$ in
\eqref{circle-cirle, agg}.
\end{remark}

We note that the evaluation of \eqref{circle-cirle, agg} requires very high computational complexity, since it involves four inseparable integrals. Therefore, we simplify \eqref{circle-cirle, agg} for the following corollaries.

\begin{corollary}[UCA within RX]\label{corollary:aggregate_mole-UCA}
When we assume that the concentration within the RX $S_0$ is uniform, the expected aggregate channel response at the RX ${\E}_{\Phi}\left\{\overline{N}_{\agg}\left(\vec{b}|\lambda\right)\right\}$ is simplified as
\begin{align}\label{circle-cirle, agg2}
{\E}_{\Phi}\left\{\overline{N}_{\agg}\left(\vec{b}|\lambda\right)\right\} \approx &\;\lambda \int_{|\vec{r}|=0}^{R_1}\int_{\varphi=0}^{2\pi}\frac{q{R_0}^2}{2D}K_0\left(\sqrt{\frac{k}{D}\Omega(\vec{b})}\right) |\vec{r}|\,d\varphi\,d|\vec{r}|.
\end{align}
\end{corollary}
\begin{corollary}[RX at Population Circle Center]\label{corollary:aggregate_mole-center}
When the RX is at the center of the circle $S_1$ where TXs are randomly distributed, the expected aggregate channel response at the RX ${\E}_{\Phi}\left\{\overline{N}_{\agg}\left(\vec{b}|\lambda\right)\right\}$ is simplified as
\begin{align}\label{circle-cirle, agg3}
&{\E}_{\Phi}\{\overline{N}_{\agg}(\vec{b}|\lambda)\}\Big{|}_{|\vec{b}|=0}\nonumber\\ 
=&\;\lambda \int_{|\vec{r}|=0}^{R_1}\int_{|\vec{r_0}|=0}^{R_0}\int_{\theta=0}^{2\pi}\frac{q}{D}K_0\left(\sqrt{\frac{k}{D}} \sqrt{|\vec{r}|^2+|\vec{r_0}|^2+2|\vec{r}||\vec{r_0}|\cos\theta}\right)|\vec{r_0}||\vec{r}|\,d\theta\,d|\vec{r_0}|\,d|\vec{r}|.
\end{align}
\end{corollary}
\begin{corollary}[RX at Population Circle Center with UCA]\label{corollary:aggregate_mole-center-UCA}
When both the concentration within the RX $S_0$ is uniform and the RX is at the center of the circle $S_1$, the expected aggregate channel response at the RX ${\E}_{\Phi}\left\{\overline{N}_{\agg}\left(\vec{b}|\lambda\right)\right\}$ is simplified as
\begin{align}\label{circle-cirle, agg4}
{\E}_{\Phi}\{\overline{N}_{\agg}(\vec{b}|\lambda)\}\Big{|}_{|\vec{b}|=0} 
\approx &\;\frac{\lambda q\pi{R_0}^2}{k}\left(1-\frac{\sqrt{k}R_1K_1\left(\sqrt{\frac{k}{D}}R_1\right)}{\sqrt{D}}\right).
\end{align}
\end{corollary}
\begin{IEEEproof}
The proofs of Corollary \ref{corollary:aggregate_mole-UCA}, Corollary \ref{corollary:aggregate_mole-center}, and
Corollary \ref{corollary:aggregate_mole-center-UCA} are given in Appendix~\ref{proof:aggregate_mole-UCA}.
\end{IEEEproof}

Although \eqref{circle-cirle, agg2} cannot be solved in the closed form, \eqref{circle-cirle, agg2} is much easier to evaluate than \eqref{circle-cirle, agg}. The numerical results in Sec. \ref{sec:Numerical} will demonstrate the accuracy of the UCA used in \eqref{circle-cirle, agg2}.

\section{Cooperating Probability at A Fixed-Location Bacterium}\label{PrCoop}
In this section, we derive the expected probability of cooperation (i.e., the number of molecules observed from itself and other PPP distributed bacteria is larger than some threshold $\eta$) at a bacterium at a fixed location $\vec{x_i}$ over the spatial random point process $\Phi(\lambda)$. We denote such a probability by $\widetilde{\prob}\left(N_{\agg}^{\dag}(\vec{x_i}|\lambda)\geq\eta\right)$. Please note that in this section  $\vec{x_i}$ is a fixed location and does not change in each instantaneous realization of the spatial random point process $\Phi(\lambda)$. In the following, we first derive the exact expression for $\widetilde{\prob}\left(N_{\agg}^{\dag}(\vec{x_i}|\lambda)\geq\eta\right)$ and then derive an approximate expression. We emphasize that deriving such a probability is challenging since the bacterial locations are different in each realization and the probability of occurrence of realizations needs to be accounted for.


%
\vspace{-2mm}
\subsection{Exact Cooperating Probability}\label{subsec:exact probability}
In this subsection, we derive the exact expression for $\widetilde{\prob}\left(N_{\agg}^{\dag}(\vec{x_i}|\lambda)\geq\eta\right)$, i.e.,
\begin{align}\label{prob,Laplace}
\widetilde{\prob}\left(N_{\agg}^{\dag}(\vec{x_i}|\lambda)\geq\eta\right)= {\E}_{\Phi}\Big\{\prob\left(N_{\agg}^{\dag}(\vec{x_i}|\lambda)\geq\eta|\overline{N}_{\agg}^{\dag}(\vec{x_i}|\lambda)\right)\Big\},
\end{align}
where ${\E}_{\Phi}$ denotes the expectation over the spatial random point process $\Phi(\lambda)$. ${N}_{\agg}^{\dag}(\vec{x_i}|\lambda)$ is the instantaneous observation at the bacterium $i$ and $\overline{N}_{\agg}^{\dag}(\vec{x_i}|\lambda)$ is its expected observation for a given instantaneous realization of random bacterial locations. \Fr{For different realizations of $\Phi(\lambda)$, $\overline{N}_{\agg}^{\dag}(\vec{x_i}|\lambda)$ changes in each realization of $\Phi(\lambda)$ since the locations of the PPP distributed bacteria change. In other words, for each realization of $\Phi$, the locations of the PPP distributed bacteria and the resulting $\overline{N}_{\agg}^{\dag}(\vec{x_i}|\lambda)$ are random. Therefore, the probability of cooperation in \eqref{prob,Laplace} is obtained by averaging the conditional probability $\prob\left(N_{\agg}^{\dag}(\vec{x_i}|\lambda)\geq\eta|\overline{N}_{\agg}^{\dag}(\vec{x_i}|\lambda)\right)$ over different realizations of $\Phi(\lambda)$ to capture this randomness.}


\Fr{We note that $N_{\agg}^{\dag}\left(\vec{x_i}|\lambda\right)$ is a Poisson binomial RV for the following reasons. We recall that $N_{\agg}^{\dag}\left(\vec{x_i}|\lambda\right)$ is the sum of $N\left(\vec{x_i}|\vec{x_j}\right)$ over $j$. We note that $N\left(\vec{x_i}|\vec{x_j}\right)$ is the sum of the number of molecules observed at the bacterium $i$ at time $t$ released from the bacterium $j$ since $t=0\,\s$. Thus, the observations at the bacterium $i$ due to continuous emission at the bacterium $j$ are not identically distributed since they are released at different times. Therefore, $N\left(\vec{x_i}|\vec{x_j}\right)$ is a Poisson binomial RV since each molecule behaves independently and has a different probability of being observed at $t=t^{\star}_i$ by the bacterium $i$ due to different releasing times. Since $N_{\agg}^{\dag}\left(\vec{x_i}|\lambda\right)$ is the sum of $N\left(\vec{x_i}|\vec{x_j}\right)$, $N_{\agg}^{\dag}\left(\vec{x_i}|\lambda\right)$ is also a Poisson binomial RV. We note that modeling $N_{\agg}^{\dag}\left(\vec{x_i}|\lambda\right)$ as a Poisson binomial RV makes the evaluation of \eqref{prob,Laplace} very cumbersome. Fortunately, in agreement with our particle-based simulation tests, $N_{\agg}^{\dag}\!\left(\vec{x_i},\infty|\lambda\right)$ can be well approximated as a Poisson RV.} Using the Poisson approximations, we rewrite \eqref{prob,Laplace} in the following lemma:
\begin{lemma}\label{lemma:coop-prob}
Assuming
${N}_{\agg}^{\dag}(\vec{x_i}|\lambda)$ is a Poisson RV with mean $\overline{N}_{\agg}^{\dag}(\vec{x_i}|\lambda)$, the cooperating probability can be written as a function of the Laplace transform of $\overline{N}_{\agg}^{\dag}(\vec{x_i}|\lambda)$. The function is given by
\begin{align}\label{prob,Laplace3}
\widetilde{\prob}\left(N_{\agg}^{\dag}(\vec{x_i}|\lambda)\geq\eta\right)
= &\;1-\sum_{n=0}^{\eta-1}\frac{1}{n!}\frac{\partial^n \mathcal{L}_{\overline{N}_{\agg}^{\dag}(\vec{x_i}|\lambda)}(-\rho)}{\partial \rho^n}\Bigg|_{\rho=-1},
\end{align}
where $\mathcal{L}_{\overline{N}_{\agg}^{\dag}(\vec{x_i}|\lambda)}(\cdot)$ is the Laplace transform of $\overline{N}_{\agg}^{\dag}(\vec{x_i}|\lambda)$, which is defined as $\mathcal{L}_{\overline{N}_{\agg}^{\dag}(\vec{x_i}|\lambda)}(s)
= {\E}_{\Phi}\Big\{\exp\left\{-s\overline{N}_{\agg}^{\dag}(\vec{x_i}|\lambda)\right\}\Big\}$.
\end{lemma}
\begin{IEEEproof}
See Appendix~\ref{proof:coop-prob}.
\end{IEEEproof}

We next derive $\mathcal{L}_{\overline{N}_{\agg}^{\dag}(\vec{x_i}|\lambda)}(s)$ in the following lemma.
\begin{lemma}\label{Theorem:Laplace}
We derive $\mathcal{L}_{\overline{N}_{\agg}^{\dag}(\vec{x_i}|\lambda)}(s)$ as
\begin{align}\label{Laplace1}
\mathcal{L}_{\overline{N}_{\agg}^{\dag}(\vec{x_i}|\lambda)}(s)
= &\;\exp\Bigg\{-s\overline{N}_{\self}-\acute{\lambda}\int_{|\vec{r}|=0}^{R_1}\int_{\varphi=0}^{2\pi}\left(1-\exp\left(-s\overline{N}(\vec{x_i}|\vec{r})\right)\right) |\vec{r}|\,d\varphi\,d|\vec{r}|\Bigg\},
\end{align}
where $\acute{\lambda}={\left(\lambda\pi R_1^2-1\right)}/{\pi R_1^2}$ and $\overline{N}(\vec{x_i}|\vec{r})$ is given by
\begin{align}\label{circle-cirle,indi-Prob2}
&\overline{N}\left(\vec{x_i}|\vec{r}\right) = \int_{|\vec{r_0}|=0}^{R_0}\int_{\theta=0}^{2\pi}\frac{q}{2D\pi}K_0\left(\sqrt{\frac{k}{D}} \Upsilon(\vec{x_i})\right)|\vec{r_0}|\,d\theta\,d|\vec{r_0}|, \quad \textrm{or} \nonumber\\
&\overline{N}\left(\vec{x_i}|\vec{r}\right) \approx \frac{q{R_0}^2}{2D}K_0\left(\sqrt{\frac{k}{D}\Omega(\vec{x_i})}\right).
\end{align}
%
\end{lemma}
\begin{IEEEproof}
See Appendix~\ref{Chap4:Proof of Laplace}.
\end{IEEEproof}

We note that the approximation one in \eqref{circle-cirle,indi-Prob2} is due to the UCA approximation within $S_0$. Using Fa\`{a} di Bruno's formula \cite{Faadi1857},  we write the $n$th derivative of $\mathcal{L}_{\overline{N}_{\agg}^{\dag}(\vec{x_i}|\lambda)}(-\rho)$ derived in \eqref{Laplace1} with respect to $\rho$ and then apply it to \eqref{prob,Laplace3}, we obtain
\begin{align}\label{prob,Laplace,fina}
\widetilde{\prob}\left(N_{\agg}^{\dag}(\vec{x_i}|\lambda)\geq\eta\right)
= &\;1-\mathcal{L}_{\overline{N}_{\agg}^{\dag}(\vec{x_i}|\lambda)}(1)\sum_{n=0}^{\eta-1}\sum\frac{1}{\prod_{j=1}^n m_j!{j!}^{m_j}}\nonumber\\
&\;\times\left(\acute{\lambda}\int_{|\vec{r}|=0}^{R_1}\int_{\varphi=0}^{2\pi}{\overline{N}(\vec{x_i}|\vec{r})}\exp\left(\rho\overline{N}(\vec{x_i}|\vec{r})\right) |\vec{r}|\,d\varphi\,d|\vec{r}|+\overline{N}_{\self}\right)^{m_1}\nonumber\\
&\;\times\prod_{j=2}^{n}\left(\acute{\lambda}\int_{|\vec{r}|=0}^{R_1}\int_{\varphi=0}^{2\pi}{\overline{N}(\vec{x_i}|\vec{r})}^{j}\exp\left(\rho\overline{N}(\vec{x_i}|\vec{r})\right) |\vec{r}|\,d\varphi\,d|\vec{r}|\right)^{m_j},
\end{align}
where the sum is over all n-tuples of nonnegative integers $(m_1,\ldots,m_n)$ satisfying the constraint $1m_1+2m_2+3m_3+\cdots+nm_n = n$.

\begin{remark}\label{remark:prob,Laplace}
The expression in \eqref{prob,Laplace,fina} comprises two integrals and thus \eqref{prob,Laplace,fina} cannot be obtained in closed form. However, \eqref{prob,Laplace,fina} can be evaluated numerically in a straightforward manner using Mathematica.%
\end{remark}

\subsection{Approximate Cooperating Probability}
In this subsection, we derive an approximate expression for $\widetilde{\prob}\left(N_{\agg}^{\dag}(\vec{x_i}|\lambda)\geq\eta\right)$ that has lower computational complexity than that of the exact expression derived in \eqref{prob,Laplace,fina}.

We recall that in \eqref{prob,Laplace}, we consider the instantaneous realization of $\overline{N}_{\agg}^{\dag}(\vec{x_i}|\lambda)$ and its PMF, which makes the evaluation of \eqref{prob,Laplace} very complicated. To ease the computational burden, we approximate the instantaneous realization of $\overline{N}_{\agg}^{\dag}(\vec{x_i}|\lambda)$ by the expected $\overline{N}_{\agg}^{\dag}(\vec{x_i}|\lambda)$ over the spatial random process $\Phi(\lambda)$, ${\E}_{\Phi}\Big\{\overline{N}_{\agg}^{\dag}(\vec{x_i}|\lambda)\Big\}$, and assume that ${N}_{\agg}^{\dag}(\vec{x_i}|\lambda)$ is a Gaussian or Poisson RV with mean ${\E}_{\Phi}\Big\{\overline{N}_{\agg}^{\dag}(\vec{x_i}|\lambda)\Big\}$. By doing so, we approximate \eqref{prob,Laplace} as:
\begin{align}\label{prob,Laplace,app}
\widetilde{\prob}\left(N_{\agg}^{\dag}(\vec{x_i}|\lambda)\geq\eta\right)
\!=\! {\E}_{\Phi}\bigg\{\!\prob\left(\!N_{\agg}^{\dag}(\vec{x_i}|\lambda)\geq\eta|\overline{N}_{\agg}^{\dag}(\vec{x_i}|\lambda)\!\right)\!\bigg\}
\!\approx\!\prob\left(\!N_{\agg}^{\dag}(\vec{x_i}|\lambda)\geq\eta|{\E}_{\Phi}\Big\{\overline{N}_{\agg}^{\dag}(\vec{x_i}|\lambda)\!\Big\}\!\right)
\end{align}

By assuming that ${N}_{\agg}^{\dag}(\vec{x_i}|\lambda)$ is a Poisson RV, we further rewrite \eqref{prob,Laplace,app} as
\begin{align}\label{prob,mean,poisson}
\widetilde{\prob}\left(N_{\agg}^{\dag}(\vec{x_i}|\lambda)\geq\eta\right)
= 1-\frac{\Gamma\left(\eta,{\E}_{\Phi}\Big\{\overline{N}_{\agg}^{\dag}(\vec{x_i}|\lambda)\Big\}\right)}{\Gamma\left(\eta\right)},
\end{align}
where ${\E}_{\Phi}\Big\{\overline{N}_{\agg}^{\dag}(\vec{x_i}|\lambda)\Big\}$ is given by
\begin{align}\label{obsExp2}
{\E}_{\Phi}\Big\{\overline{N}_{\agg}^{\dag}(\vec{x_i}|\lambda)\Big\}
= &\;{\E}_{\Phi}\Bigg\{\sum_{\vec{x_j}\in\Phi\left(\lambda\right)}\overline{N}\left(\vec{x_i}|\vec{x_j}\right)\Bigg\}
= {\E}_{\Phi}\Bigg\{\overline{N}\left(\vec{x_i}|\vec{x_i}\right)+\sum_{\vec{x_j}\in\Phi\left(\lambda\right)/\vec{x_i}}\overline{N}\left(\vec{x_i}|\vec{x_j}\right)\Bigg\}\nonumber\\
= &\;\overline{N}_{\self}+{\E}_{\Phi}\Bigg\{\sum_{\vec{a}\in\Phi(\acute{\lambda})}\overline{N}(\vec{x_i}|\vec{a})\Bigg\}
= \overline{N}_{\self}+{\E}_{\Phi}\bigg\{\overline{N}_{\agg}(\vec{x_i}|\acute{\lambda})\bigg\},
\end{align}
where $\E\{\overline{N}_{\agg}(\vec{x_i}|\acute{\lambda})\}$ can be obtained by replacing $|\vec{b}|$ with $|\vec{x_i}|$ and $\lambda$ with $\acute{\lambda}$ in \eqref{circle-cirle, agg} if the UCA is not valid or \eqref{circle-cirle, agg2} if the UCA is valid.

\section{Characterization of Number of Cooperative Bacteria}\label{cooperatorNumber}
In this section, we characterize the distribution of the number of cooperators. To this end, we first derive the MGF of the number of cooperators. Using the derived MGF, we then derive the expressions for the moments and cumulants of the number of cooperators. Using the derived moments and cumulants, we study the convergence of the distribution of the number of cooperators to a Gaussian distribution. Furthermore, we derive the expected number of pairs of two nearest bacteria both cooperating, which can be used to study the impact of a cooperative bacterium on the behaviors of the neighboring bacteria in a QS system. The problem addressed in this section is challenging since we need to consider the random received signal at each bacterium in a random location due to the random motion of molecules released from a population of randomly-distributed bacteria.


\subsection{Moment and Cumulant Generating Functions}
We denote the decision of cooperation and noncooperation of the bacterium $i$ by $B(\vec{x_i},\Phi)=1$ and $B(\vec{x_i},\Phi)=0$, respectively. We note that $B(\vec{x_i},\Phi)$ is a Bernoulli RV with mean $\overline{B}(\vec{x_i},\Phi)$. We denote the number of all cooperators by $Z$, i.e., $Z=\sum_{\vec{x_i}\in\Phi(\lambda)}B(\vec{x_i},\Phi)$. We first derive the exact expression for the MGF of $Z$ and then provide an approximated expression that can be readily used to derive the $n$th moment and the $n$th cumulant of $Z$. We derive the exact expression for the MGF of $Z$ in the following theorem.
\begin{theorem}\label{theorem:MGF-exact}
The exact expression for the MGF of $Z$, ${M}_{Z}(u)$, is given by
\begin{align}\label{mgf3}
{M}_{Z}(u)=&\;{\E}_{\Phi}\Bigg\{\prod_{\vec{x_i}\in\Phi(\lambda)}h(\vec{x_i},\Phi)\Bigg\},
\end{align}
where $h(\vec{x_i},\Phi)$ is given by
\begin{align}\label{h}
h(\vec{x_i},\Phi)=&\;1+(\exp(u)-1)\!\left(\!1-\left(\!\sum_{n=0}^{\eta-1}\frac{1}{n!}\exp\!\Bigg\{\!-\sum_{\vec{x_j}\in\Phi(\lambda)}\overline{N}(\vec{x_i}|\vec{x_j})\!\Bigg\}\!\left(\!\sum_{\vec{x_j}\in\Phi(\lambda)}\overline{N}(\vec{x_i}|\vec{x_j})\!\right)^{n}\!\right)\!\right). \end{align}
\end{theorem}
\begin{IEEEproof}
See Appendix \ref{proof:MGF-exact}.
\end{IEEEproof}

We note that $h(\vec{x_i},\Phi)$ not only depends on $\vec{x_i}$ but also depends on the locations of the other bacteria in $\Phi$. Hence, it is mathematically intractable to write $\E\{\prod_{x\in\Phi} h(x,\Phi)\}$ as a form that only includes addition, multiplication, or integrals using existing tools in stochastic geometry, which makes deriving moments or cumulants based on \eqref{mgf3} cumbersome. To tackle this problem, we next derive an approximated expression for ${M}_{Z}(u)$ in the following theorem.
\begin{theorem}\label{theorem:MGF-approximate}
The approximated expression for ${M}_{Z}(u)$ is given by
\begin{align}\label{mgf, app1}
{M}_{Z}(u)\approx&\; \exp\left(-\lambda\int_{|\vec{r_1}|=0}^{R_1}(1-\exp(u))\widetilde{\prob}\left(N_{\agg}^{\dag}(\vec{r_1}|\lambda)\geq\eta\right) 2\pi|\vec{r_1}|d|\vec{r_1}|\right),
\end{align}
where $\widetilde{\prob}\left(N_{\agg}^{\dag}(\vec{r_1}|\lambda)\geq\eta\right)$ can be obtained by replacing $\vec{x_i}$ with $\vec{r_1}$ in \eqref{prob,Laplace,fina} or \eqref{prob,mean,poisson}.
We also derive the approximated cumulant generating function of $Z$, ${\mathcal{K}}_{Z}(u)$, as
\begin{align}\label{cgf}
{\mathcal{K}}_{Z}(u)
\approx-\lambda\int_{|\vec{r_1}|=0}^{R_1}(1-\exp(u))\widetilde{\prob}\left(N_{\agg}^{\dag}(\vec{r_1}|\lambda)\geq\eta\right)2\pi|\vec{r_1}|d|\vec{r_1}|.
\end{align}
\end{theorem}
\begin{IEEEproof}
See Appendix \ref{proof:approximate}.
\end{IEEEproof}

We discuss the accuracy of the approximation in \eqref{mgf, app1} and \eqref{cgf} in the following remark:
\begin{remark}\label{accuracy}
Based on Appendix \ref{proof:approximate}, the approximation in \eqref{mgf, app1} and \eqref{cgf} is because we use the expected cooperating probability over the spatial point process $\Phi$ to approximate the conditional cooperating probability for a given instantaneous realization of this point process $\Phi$. Intuitively, this approximation is more accurate when the density of the bacterial population, $\lambda$, is lower. This is because when the density is lower,
the instantaneous number of received molecules from other bacteria is closer to the expected number of received molecules over the spatial point process $\Phi$. Hence, the cooperating probability for a given instantaneous realization of $\Phi$ is closer to that expected over $\Phi$, thus the approximation is more accurate. The numerical results in Sec. \ref{sec:Numerical} will verify this intuition. 
\end{remark}

\subsection{Moments, Cumulants, and Distribution}\label{subsec:mo}
We derive the $n$th moment of $Z$ and the $n$th cumulant of $Z$ in the following theorem.
\begin{theorem}\label{theorem:moment,cumulant}
The expression for the $n$th moment of $Z$ is given by
\begin{align}\label{moment,fina}
{\E}_{\Phi}\{(Z)^n\}\approx&\;\sum\frac{n!}{\prod_{j=1}^n m_j!{j!}^{m_j}}\prod_{j=1}^{n}\left(\lambda\int_{|\vec{r_1}|=0}^{R_1}\widetilde{\prob}\left(N_{\agg}^{\dag}(\vec{r_1}|\lambda)\geq\eta\right) 2\pi|\vec{r_1}|d|\vec{r_1}|\right)^{m_j},
\end{align}
where the sum is over all n-tuples of nonnegative integers $(m_1,\ldots,m_n)$ satisfying the constraint $1m_1+2m_2+3m_3+\cdots+nm_n = n$. The expression for the $n$th cumulant of $Z$, denoted by $\kappa(n)$, is given by
\begin{align}\label{cumulant}
\kappa(n)
\approx\lambda\int_{|\vec{r_1}|=0}^{R_1}\widetilde{\prob}\left(N_{\agg}^{\dag}(\vec{r_1}|\lambda)\geq\eta\right)2\pi|\vec{r_1}|d|\vec{r_1}|.
\end{align}
\end{theorem}
\begin{IEEEproof}
\eqref{moment,fina} can be obtained by ${\E}_{\Phi}\{(Z)^n\} = \frac{\partial^n{M}_{Z}(u)}{\partial u^n}\bigg{|}_{u=0}$ and Fa\`{a} di Bruno's formula. \eqref{cumulant} can be obtained by $\kappa(n) = \frac{\partial^n{\mathcal{K}}_{Z}(u)}{\partial u^n}|_{u=0}$ and the expression for $\mathcal{K}_{Z}(u)$ derived in \eqref{cgf}.
\end{IEEEproof}

Based on \eqref{moment,fina}, we have the following propositions about the moments of $Z$:
\begin{proposition}\label{mean1}
The approximation of the first moment of $Z$, ${\E}_{\Phi}\{Z\}$, given by \eqref{moment,fina} is tight, i.e.,
\begin{align}\label{mean,fina}
{\E}_{\Phi}\{Z\}=&\;\lambda\int_{|\vec{r_1}|=0}^{R_1}\widetilde{\prob}\left(N_{\agg}^{\dag}(\vec{r_1}|\lambda)\geq\eta\right) 2\pi|\vec{r_1}|d|\vec{r_1}|.
\end{align}
\begin{IEEEproof}
See Appendix \ref{proof:mean1}.
\end{IEEEproof}
\end{proposition}
\begin{proposition}\label{remark:mean}
When the density of the bacterial population, $\lambda$, is relatively low, the variance of $Z$, denoted by $\var\{Z\}$, can be well approximated by its mean ${\E}_{\Phi}\{{Z}\}$, i.e.,
\begin{align}\label{var}
\var\{Z\}\approx{\E}_{\Phi}\{Z\}.
\end{align}
\begin{IEEEproof}
Applying $n=2$ to \eqref{moment,fina} and combining with \eqref{mean,fina}, we obtain the second moment of $Z$ as
${\E}_{\Phi}\{(Z)^2\}\approx\left({\E}_{\Phi}\{{Z}\}\right)^2+{\E}_{\Phi}\{{Z}\}$.
Using this relation and $\var\{Z\} = {\E}_{\Phi}\{(Z)^2\}-({\E}_{\Phi}\{Z\})^2$, we obtain $\var\{Z\}\approx{\E}_{\Phi}\{Z\}$. In addition, as discussed in Remark~\ref{accuracy}, the approximation used in \eqref{mgf, app1} and \eqref{cgf} is more accurate when the density $\lambda$ is lower. This complete the proof.
\end{IEEEproof}
\end{proposition}

Interestingly, by combining \eqref{moment,fina}, \eqref{mean,fina}, and \eqref{cumulant}, we obtain the relation between ${\E}_{\Phi}\{(Z)^n\}$, ${\E}_{\Phi}\{Z\}$, and $\kappa(n)$, as follows:
\begin{align}\label{moment,relation,1}
{\E}_{\Phi}\{(Z)^n\}\approx\sum\frac{n!}{\prod_{j=1}^n m_j!{j!}^{m_j}}\prod_{j=1}^{n}\left({\E}_{\Phi}\{Z\}\right)^{m_j},
\quad\kappa(n)\approx{\E}_{\Phi}\{Z\}.
\end{align}
%

Thus, once ${\E}_{\Phi}\{Z\}$ in \eqref{mean,fina} is determined, ${\E}_{\Phi}\{(Z)^n\}$ and $\kappa(n)$ can be easily determined via \eqref{moment,relation,1}. 

We discuss the effect of motion of bacteria on the number of cooperators in the following remark:
\begin{remark}\label{remark:move}
\Fr{If we assume bacteria experience unbiased diffusive motion over time after being randomly placed, our simulation results (see Appendix \ref{proof:move}) show that the number of cooperators would decrease as the bacteria's diffusion coefficient increases (i.e., as bacteria move faster). This is because when bacteria move faster, the distance between bacteria is larger and they observe fewer molecules, which leads to a smaller cooperation probability and fewer cooperators.}
\end{remark}


We finally investigate the distribution of the number of cooperators, $Z$. The skewness and kurtosis describe the symmetry and peakedness of the distribution of a RV, respectively. Using \eqref{cumulant} and \cite{Hogg}, we derive the skewness, $\beta_1$ and kurtosis, $\beta_2$, of $Z$ as
\begin{align}\label{skewness,kurtosis}
\beta_1 =\frac{\kappa(3)}{\kappa(2)^{3/2}}\approx\;\left({\E}_{\Phi}\{{Z}\}\right)^{-\frac{1}{2}}, \quad \beta_2 = \;\frac{\kappa(4)}{\kappa(2)^{2}}\approx\;\left({\E}_{\Phi}\{{Z}\}\right)^{-1}.
\end{align}
Based on \cite{DeCarlo}, the skewness and kurtosis together can be employed to assess the normality of a distribution. For a Gaussian distribution, $\beta_1=\beta_2=0$. Thus, if both $\beta_1\rightarrow0$ and $\beta_2\rightarrow0$, we can say that the RV could be closely approximated by a Gaussian distribution \cite{Haenggi2007}. Based on \eqref{skewness,kurtosis}, we see that $Z$ can be approximated by a Gaussian distribution when ${\E}_{\Phi}\{{Z}\}\rightarrow\infty$ since ${\E}_{\Phi}\{{Z}\}\rightarrow\infty$ leads to $\beta_1\rightarrow0$ and $\beta_2\rightarrow0$. Using ${\E}_{\Phi}\{{Z}\}$ and $\var\{Z\}$ derived in this subsection, we can use well-known closed-form distributions (e.g., Poisson and Gaussian distributions) to approximate the PMF and CDF of $Z$. In Sec. \ref{sec:Numerical}, we will use Poisson and Gaussian distributions with derived mean and variance to fit the PMF and CDF of the number of cooperators.

\subsection{Pairs of Two Nearest Bacteria Both Cooperating}
\Fr{Using the cooperating probability derived in Section IV, we evaluate the expected number of pairs of one node and its $n$th nearest node to both be cooperators, which we denote by $P(n)$. Such an expression can be used to quantify the average number of clusters of cooperators and to study the impact of a cooperative bacterium on the cooperative behaviors of its neighbors. For example, we could compare the number of pairs of two nearest bacteria both cooperating and the number of cooperators derived in Section V.B to know whether most cooperators are the only cooperators in their vicinity, which shows the impact of a cooperative bacterium on the behaviors of its neighbors.} We first write $P(n)$ as
\begin{align}\label{Prnear}
P(n)={\E}_{\Phi}\Bigg\{\sum_{\vec{x_i}\in\Phi(\lambda)}\{\prob\left(B(\vec{x_i},\Phi)=1\right)\prob\left(B(\vec{x}_{i,n},\Phi)=1\right)\}\Bigg\},
\end{align}
where $\vec{x}_{i,n}$ is the $n$th nearest node to node $\vec{x_i}$. We derive $P(n)$ in the following theorem.
\begin{theorem}\label{theorem:nearest}
The expression of $P(n)$ in given by
\begin{align}\label{near3}
P(n)
=&\;\lambda\int_{|\vec{r_1}|=0}^{R_1}\bigg\{\widetilde{\prob}\left(N_{\agg}^{\dag}(\vec{r_1}|\lambda)\geq\eta\right)\int_{|\vec{r_2}|=0}^{R_1}\int_{\psi=0}^{2\pi}\widetilde{\prob}\left(N_{\agg}^{\dag}(\vec{r_2}|\lambda)\geq\eta\right)
\frac{g_{n}(r(\vec{r_1}))}{2\pi r(\vec{r_1})}|\vec{r_2}|\,d|\vec{r_2}|\,d\psi\bigg\}2\pi|\vec{r_1}|\,d|\vec{r_1}|,
\end{align}
where $\widetilde{\prob}\left(N_{\agg}^{\dag}(\vec{x}|\lambda)\geq\eta\right)$ can by obtained by replacing $\vec{x_i}$ with $\vec{x}$ in \eqref{prob,Laplace,fina} or \eqref{prob,mean,poisson}.
\end{theorem}
\begin{IEEEproof}
See Appendix \ref{proof:nearest}.
\end{IEEEproof}


\vspace{-4mm}
\section{Numerical Results and Simulations}\label{sec:Numerical}

In this section, we present simulation and numerical results to assess the accuracy of our derived analytical results and reveal the impact of environmental parameters on the number of molecules observed, the cooperating probability, and the statistics of the number of cooperators derived in Sections \ref{sec:2D}--\ref{cooperatorNumber}.

\begin{figure}[!t]
\centering
\subfigure[Impulse Emission]{\label{fig-1b}\includegraphics[height=2.4in]{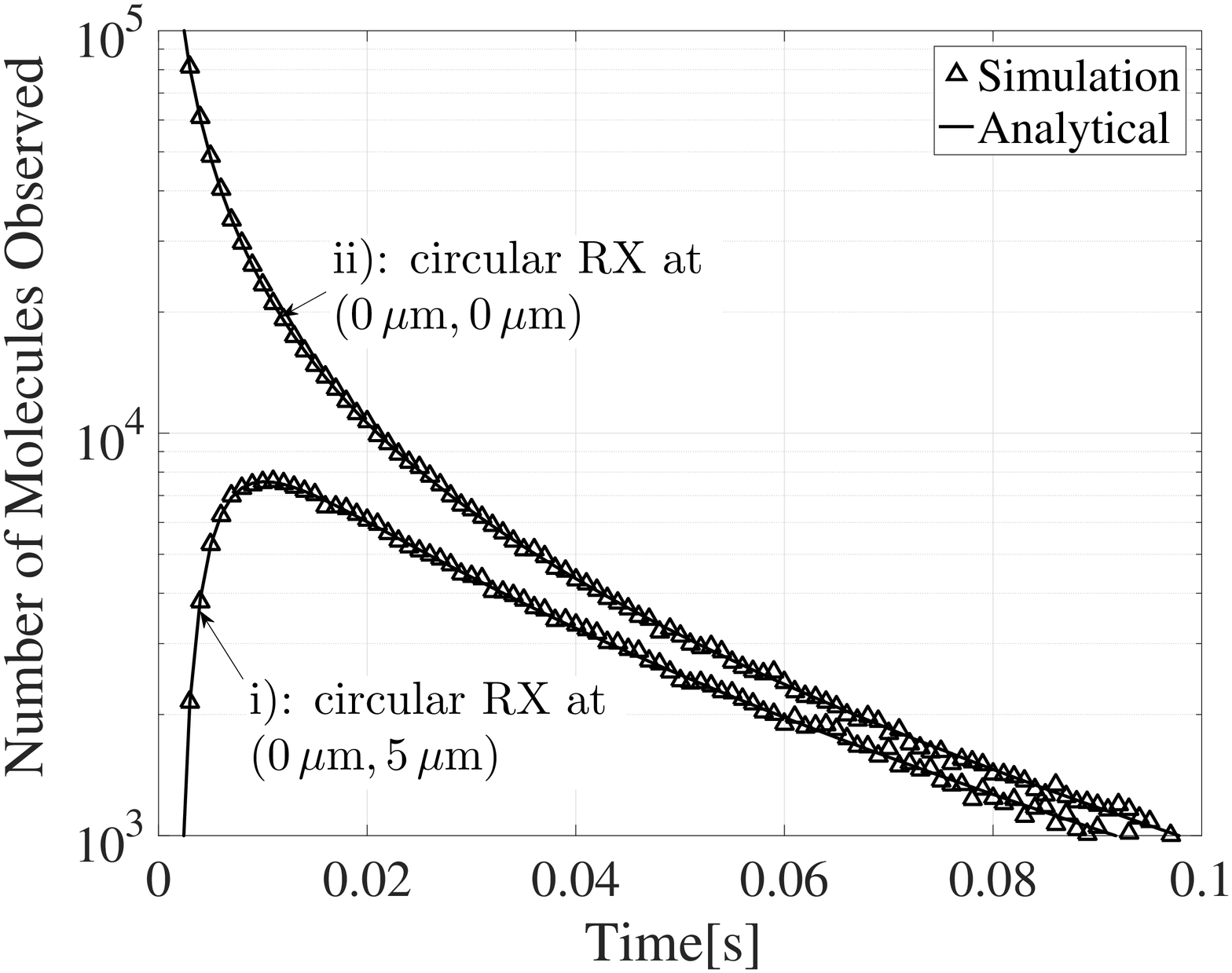}}
\subfigure[Continuous Emission]{\label{Fig-1a}\includegraphics[height=2.4in]{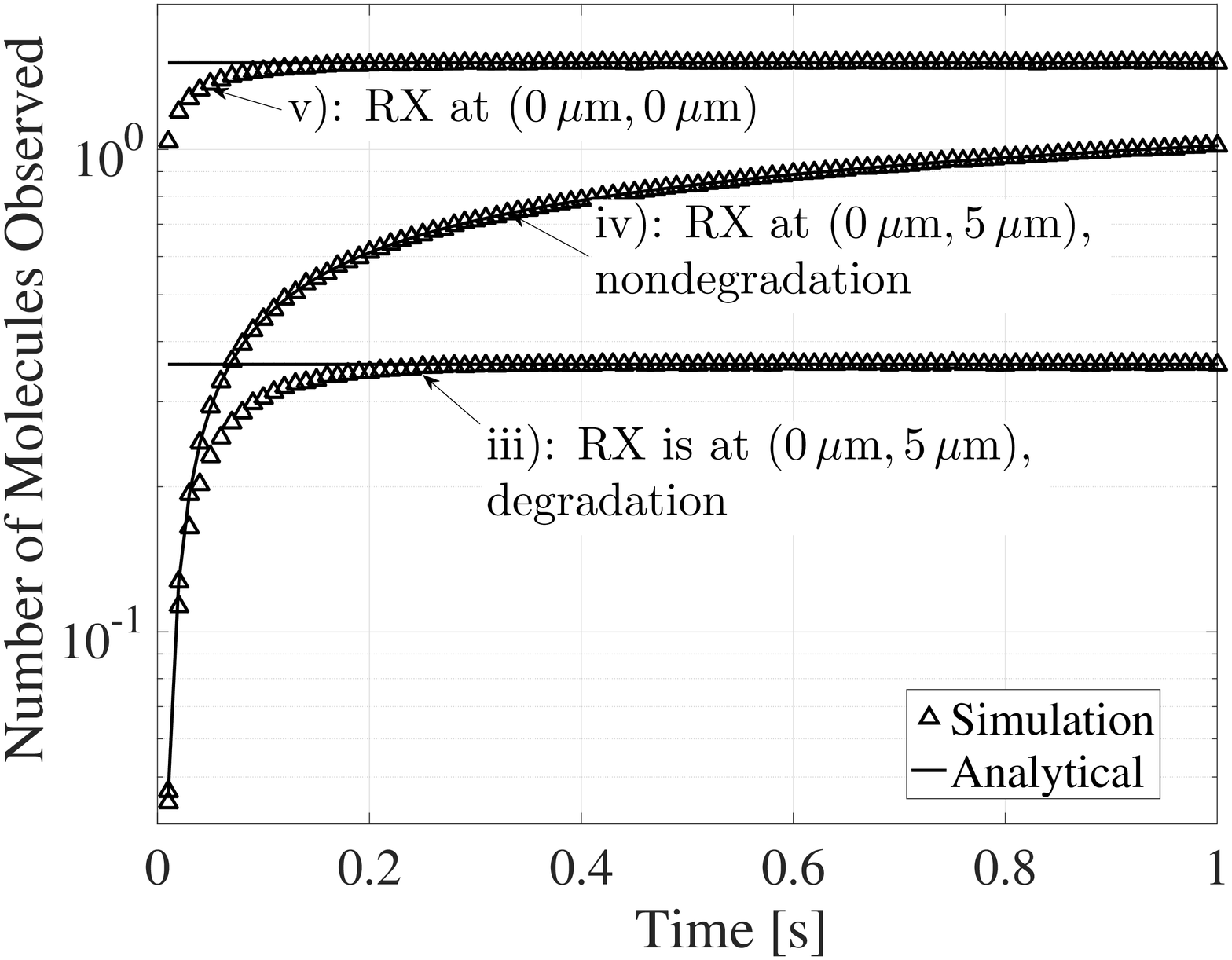}}
\vspace{-2mm}
\caption{The expected number of molecules observed at the RX $\overline{N}\left(\vec{b},t\right)$ versus time due to the emission of one TX located at $(0,0)$. In Fig. \ref{fig-1b}, we consider one \emph{impulse emission} with $10^5$ molecules and molecular degradation is considered. We consider two cases of the RX in Fig. \ref{fig-1b}: Case i) the circular RX located at $(0, 5\,{\mu}\m)$ and Case ii) the circular RX located at $(0,0)$. In Fig. \ref{Fig-1a}, we consider continuous emission and the circular RX is considered. We consider three cases of the RX in Fig. \ref{Fig-1a}: Case iii) the RX located at $(0, 5\,{\mu}\m)$ with molecular degradation, Case iv) the RX located at $(5\,{\mu}\m,0)$ without molecular degradation, and Case v) the RX located at $(0,0)$ with molecular degradation.}
\label{fig-1}
\vspace{-6mm}
\end{figure}

\begin{figure}[!t]
\centering
\subfigure[The Expected Number of Molecules Observed]{\label{fig-1c}\includegraphics[height=2.4in]{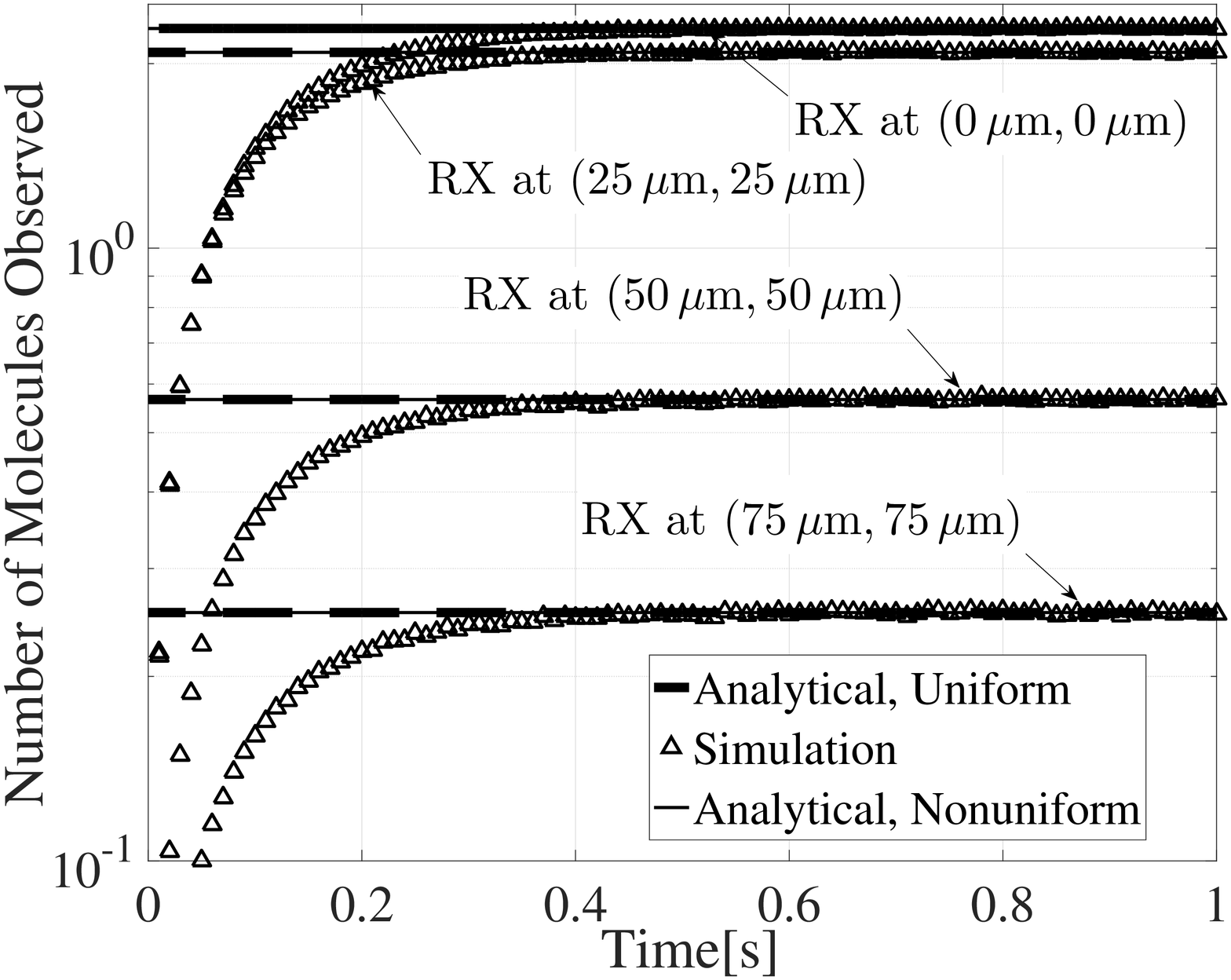}}
\subfigure[The Cooperating Probability]{\label{Fig-2a}\includegraphics[height=2.4in]{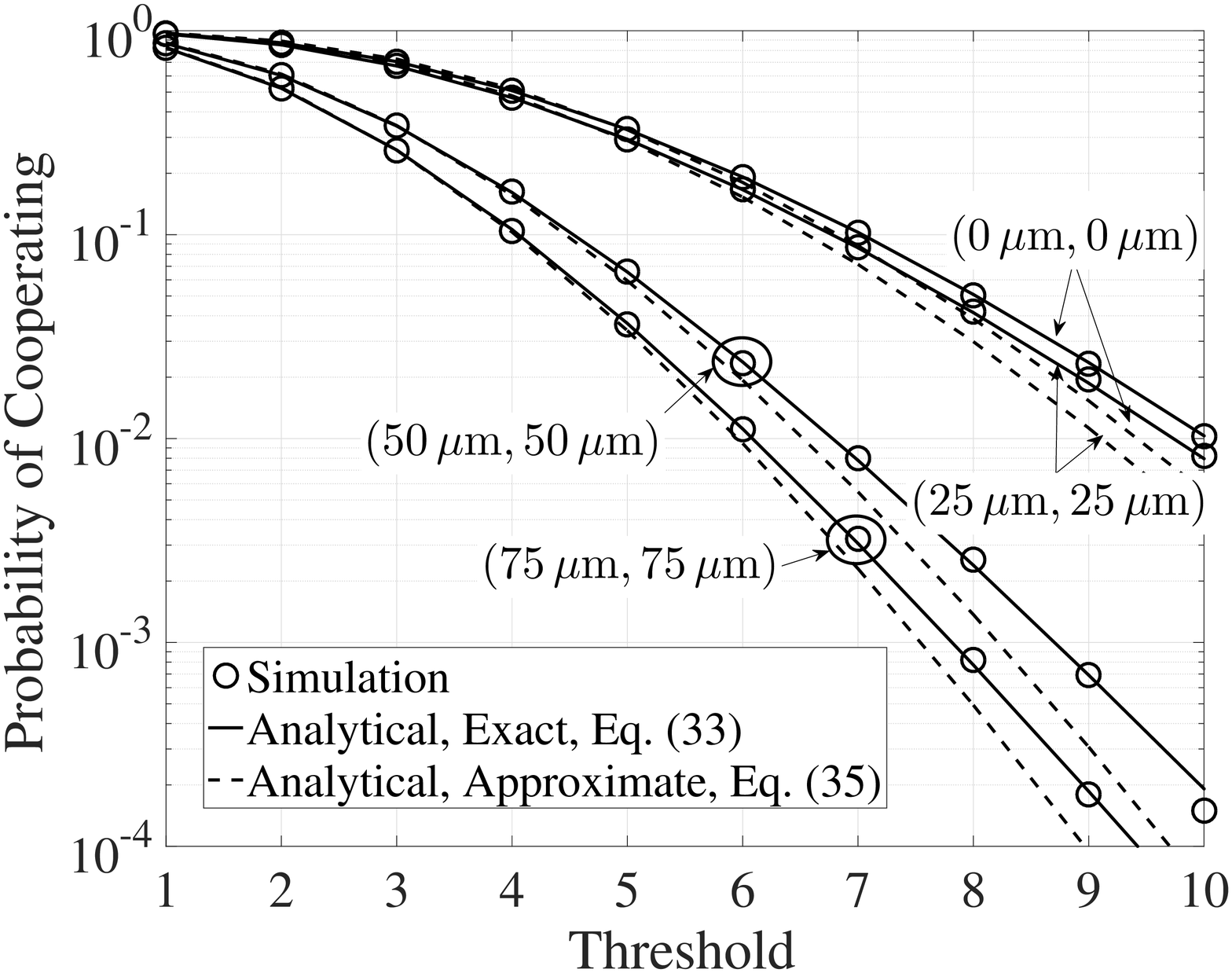}}
\vspace{-2mm}
\caption{The expected number of molecules observed at the RX, ${\E}_{\Phi}\left\{\overline{N}_{\agg}\left(\vec{b}|\lambda\right)\right\}$, in Fig. \ref{fig-1c} and the corresponding cooperating probability at the RX, $\widetilde{\prob}\left(N_{\agg}^{\dag}(\vec{x_i}|\lambda)\geq\eta\right)$, in Fig. \ref{Fig-2a} due to continuous emission at randomly-distributed TXs. For different environmental radii $R_1=50\,{\mu}\m$, $R_1=100\,{\mu}\m$, and $R_1=150\,{\mu}\m$, the RX's location is $(\frac{R_1}{2},\frac{R_1}{2})$. For $R_1=50\,{\mu}\m$, we also consider the RX located at the center of environment, i.e., $(0,0)$.}\label{fig-2}
\vspace{-6mm}
\end{figure}

\begin{figure*}[!tbp]
						\centering
						\begin{minipage}[t]{0.49\textwidth}\hspace*{-5 mm}
							\centering
							\resizebox{1.05\linewidth}{!}{
								\includegraphics[scale=0.55]{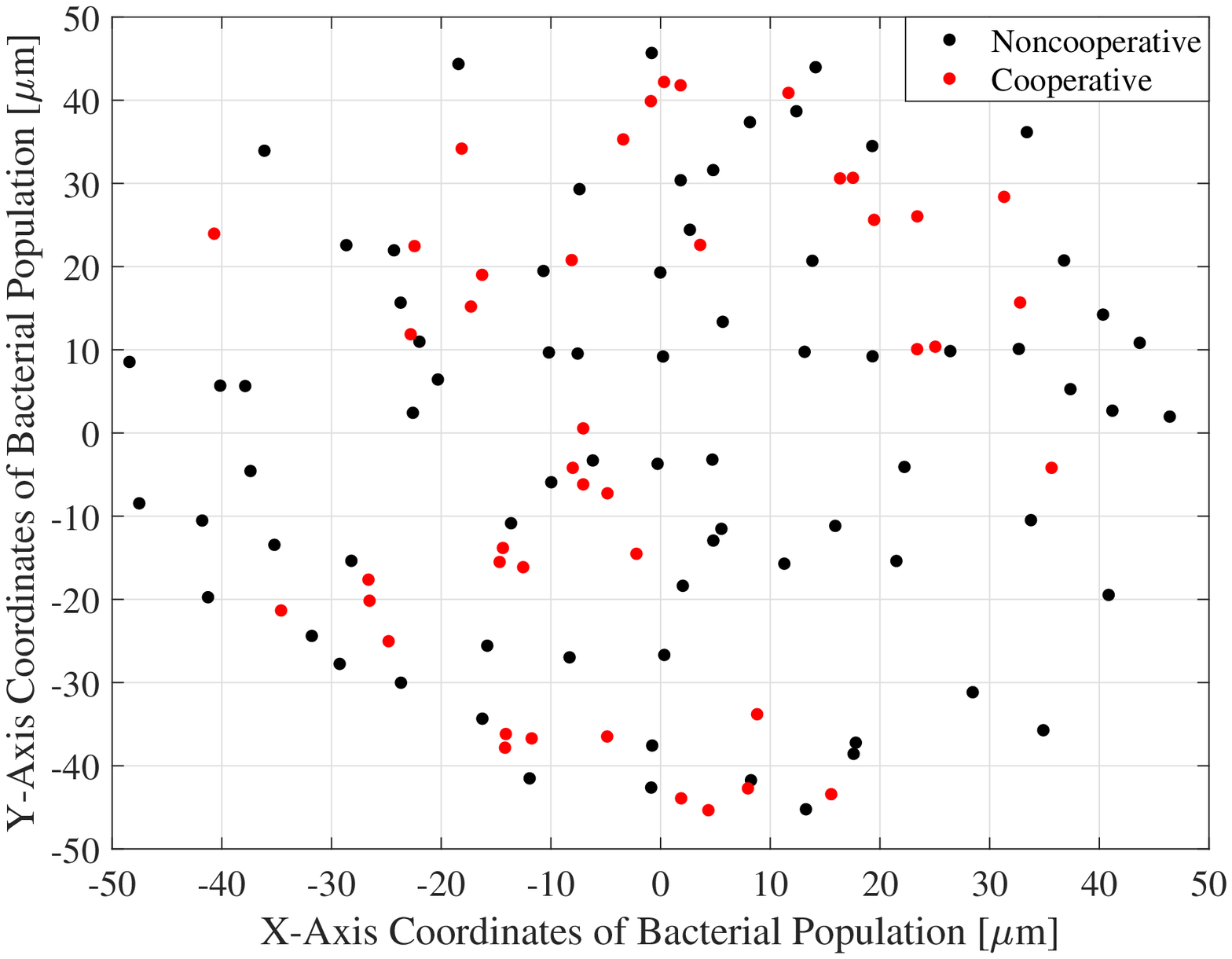}}\vspace*{-4 mm}
							\caption{
								\Fr{A realization of a 2D PPP of randomly-distributed bacteria and the resulting cooperative bacteria. $R_1 = 50\,{\mu}\m$.}}	
							\label{fig:spatial}
						\end{minipage}
						\hfill
						\begin{minipage}[t]{0.1\textwidth}
						\end{minipage}
						\vspace*{-1 mm}
						\begin{minipage}[t]{0.49\textwidth}\hspace*{-5 mm}
							\centering
							\resizebox{1.05\linewidth}{!}{
								\includegraphics[scale=0.55]{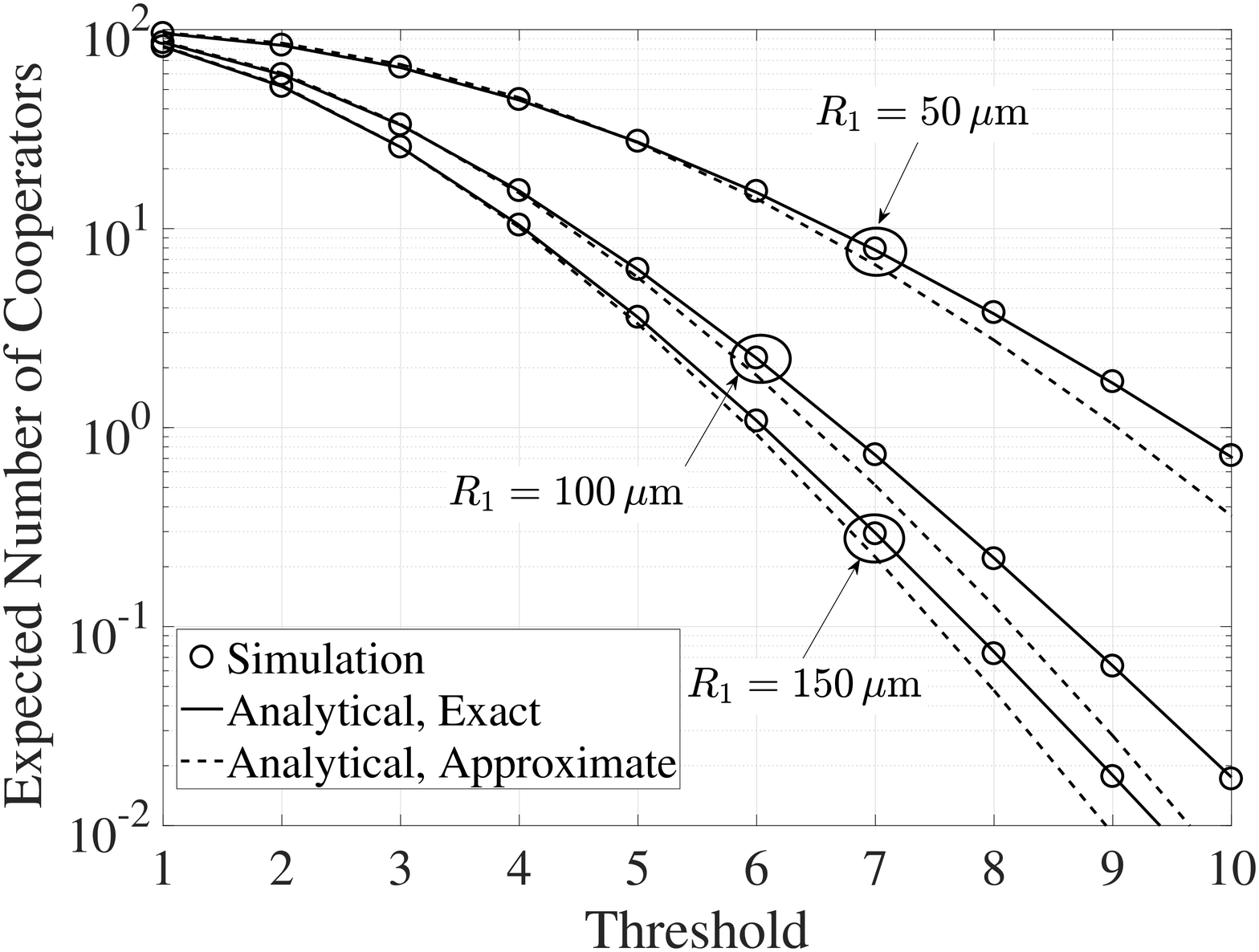}}\vspace*{-4 mm}
							\caption{
								The expected number of cooperators over spatial PPP ${\E}_{\Phi}\{Z\}$ versus threshold $\eta$ for different population radii $R_1$. 	}	
							\label{Fig-2b}
						\end{minipage}
\vspace{-6mm}
					\end{figure*}

\begin{figure}[!t]
\centering
\subfigure[Moments:~$R_1=50\,{\mu}\m$]{\label{fig3:a}\includegraphics[height=1.75in]{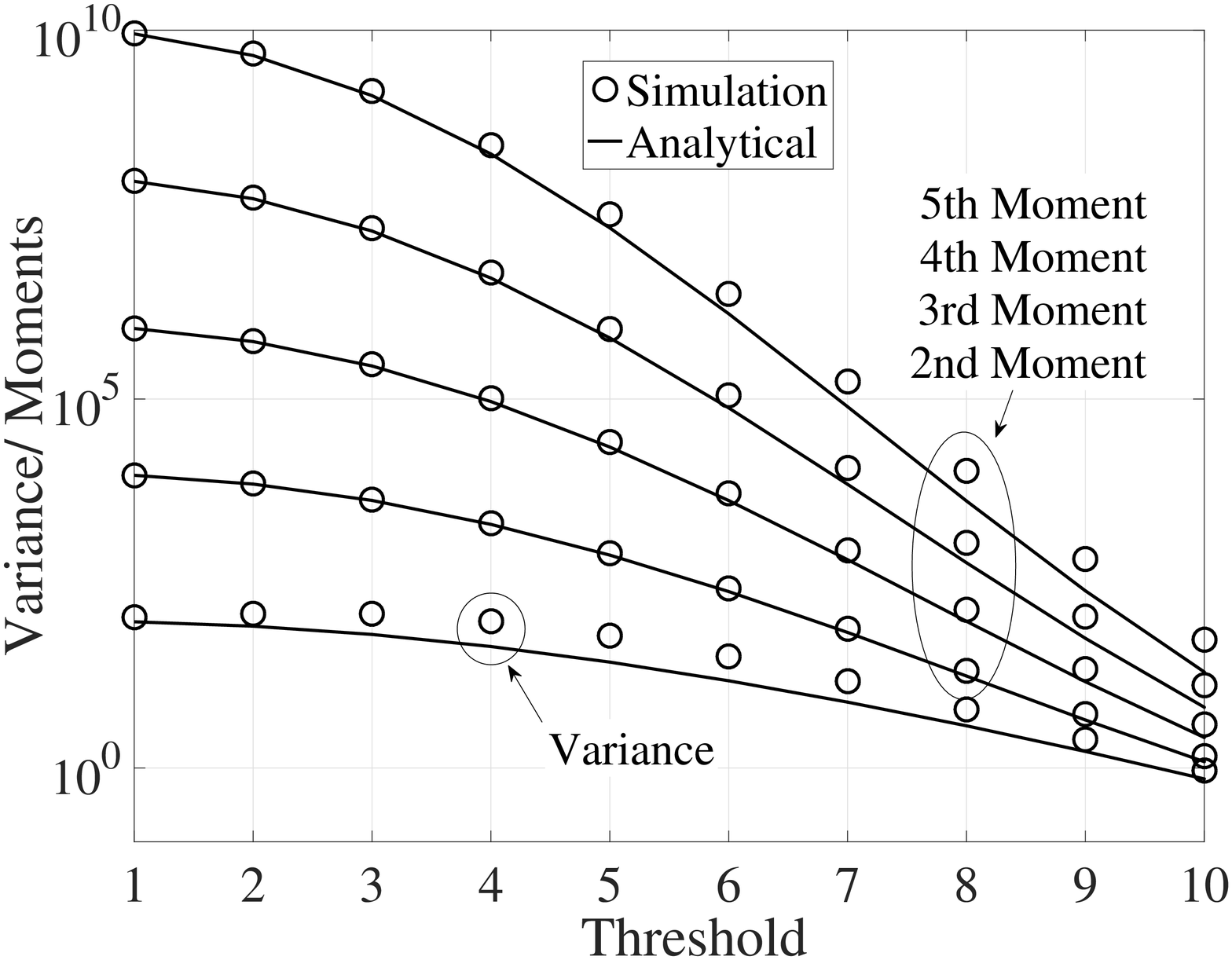}}
\subfigure[Moments:~$R_1=100\,{\mu}\m$]{\label{fig3:b}\includegraphics[height=1.75in]{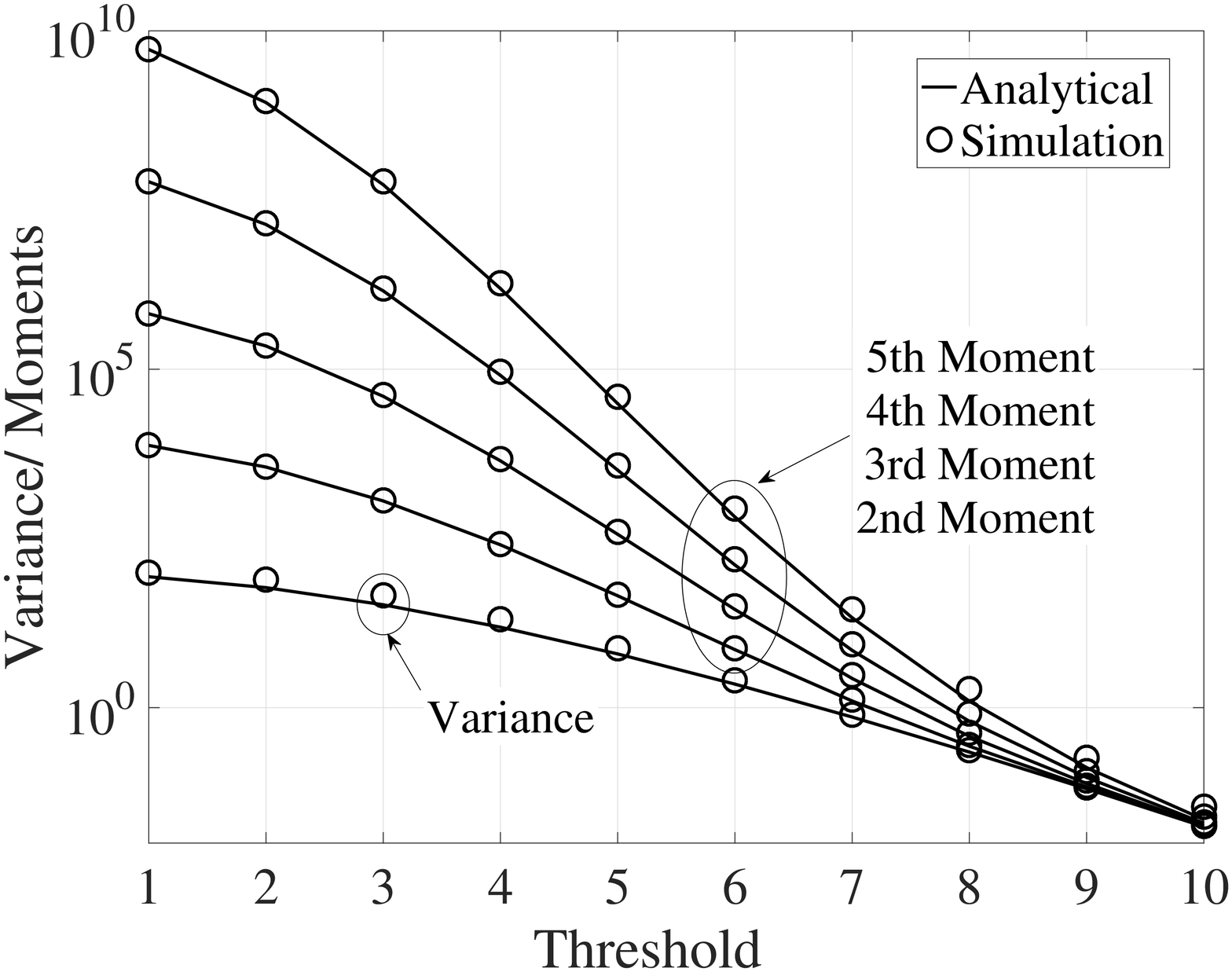}}
\subfigure[Moments:~$R_1=150\,{\mu}\m$]{\label{fig3:c}\includegraphics[height=1.75in]{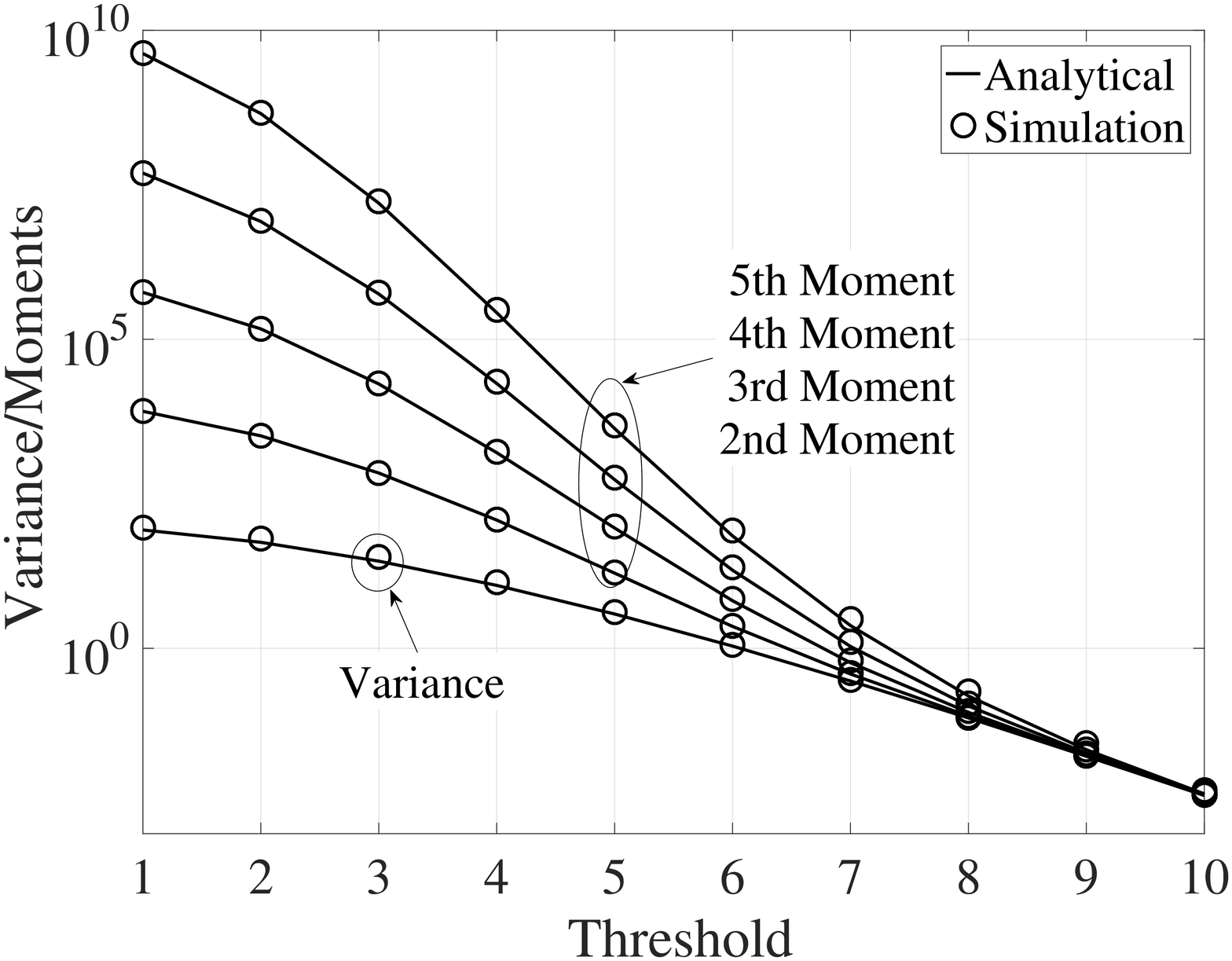}}
\caption{The different orders of moments of number of cooperators ${\E}_{\Phi}\{(Z)^n\}$ versus threshold $\eta$ for different population radii $R_1$.}\label{fig3}
\vspace{-6mm}
\end{figure}

\begin{figure}[!t]
\centering
\subfigure[$R_1=50\,{\mu}\m$, $\eta=1$]{\label{fig4:1}\includegraphics[height=1.6in]{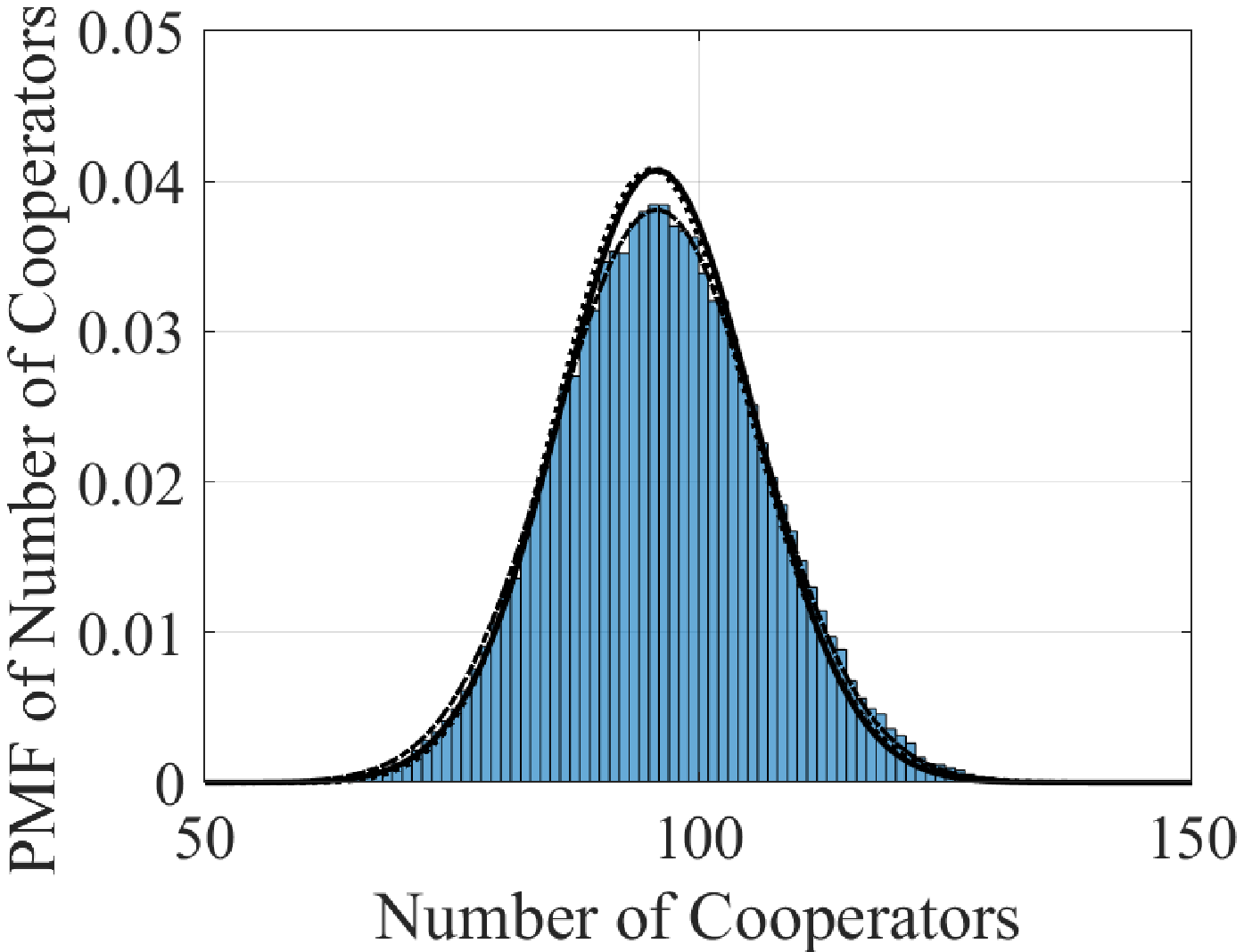}}
\subfigure[$R_1=100\,{\mu}\m$, $\eta=1$]{\label{fig4:2}\includegraphics[height=1.6in]{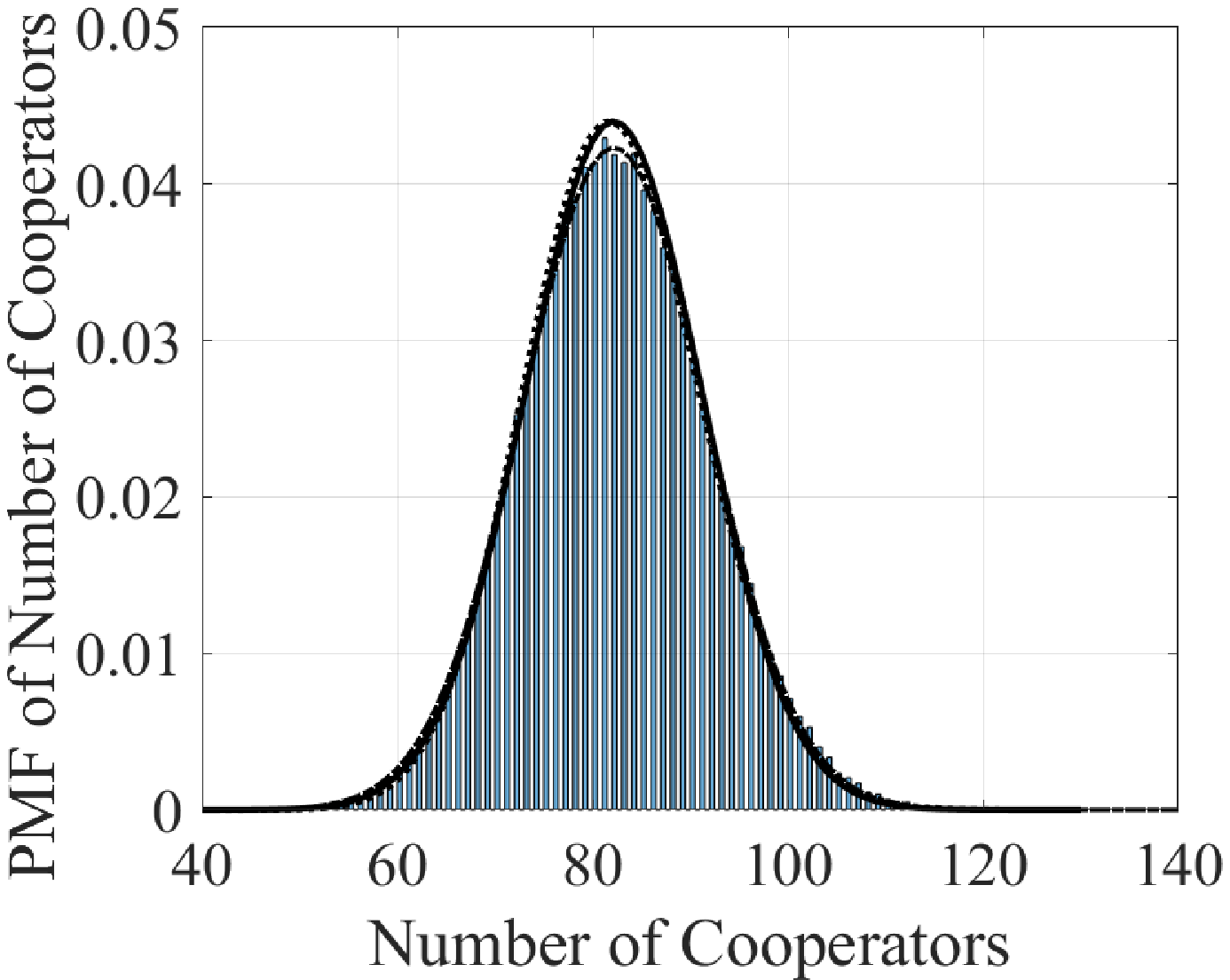}}
\subfigure[$R_1=150\,{\mu}\m$, $\eta=1$]{\label{fig4:3}\includegraphics[height=1.6in]{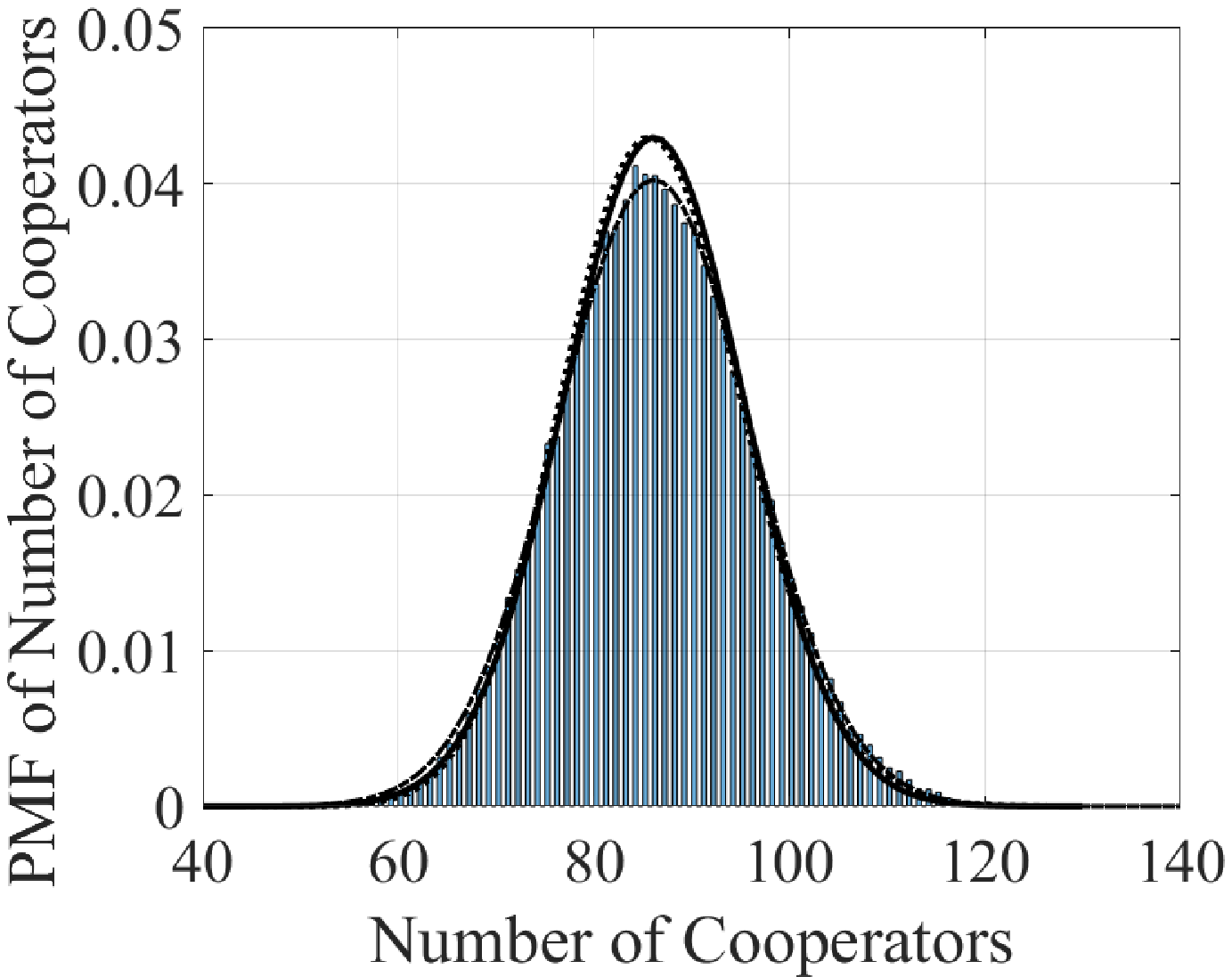}}
\subfigure[$R_1=50\,{\mu}\m$, $\eta=5$]{\label{fig4:4}\includegraphics[height=1.6in]{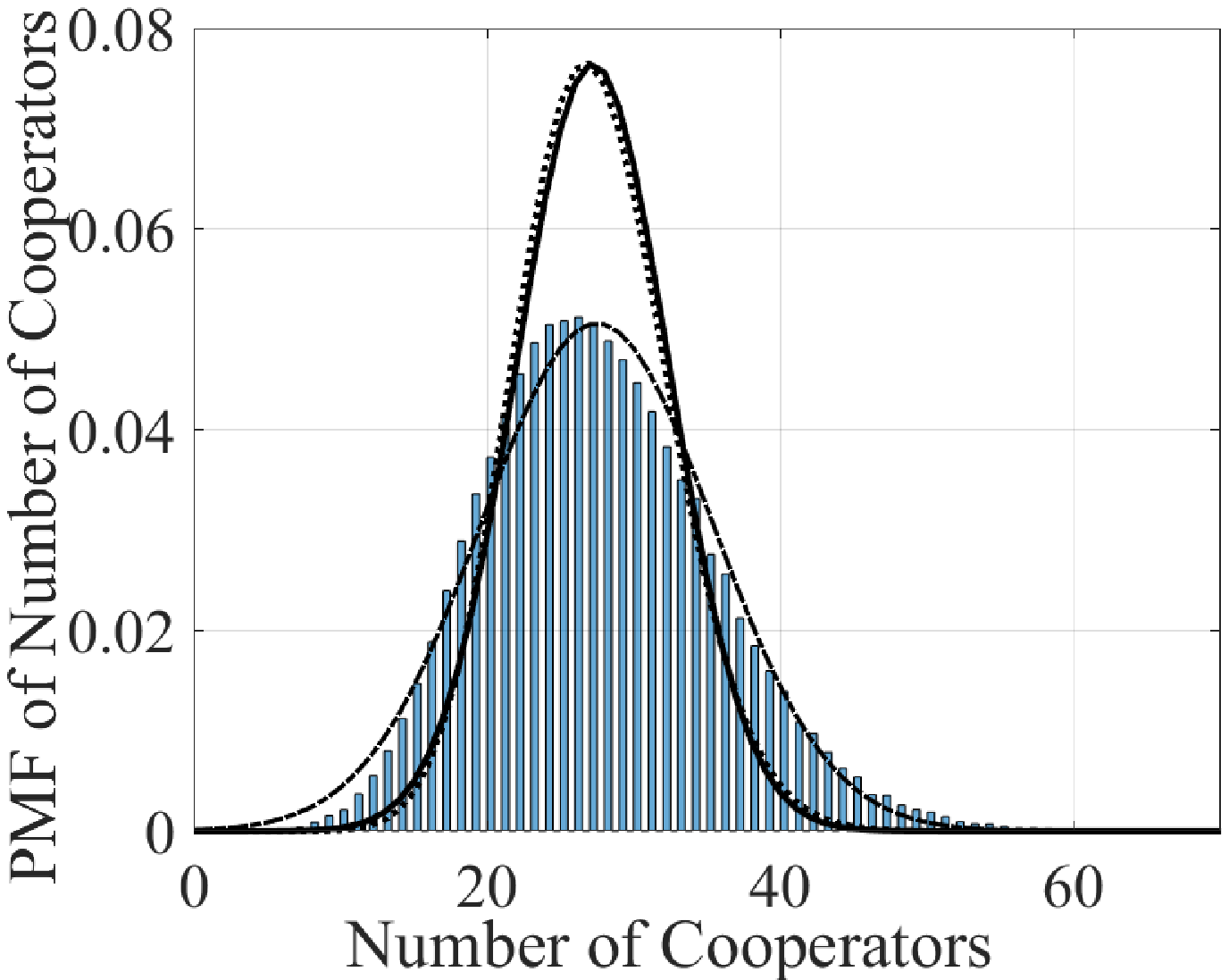}}
\subfigure[$R_1=100\,{\mu}\m$, $\eta=5$]{\label{fig4:5}\includegraphics[height=1.6in]{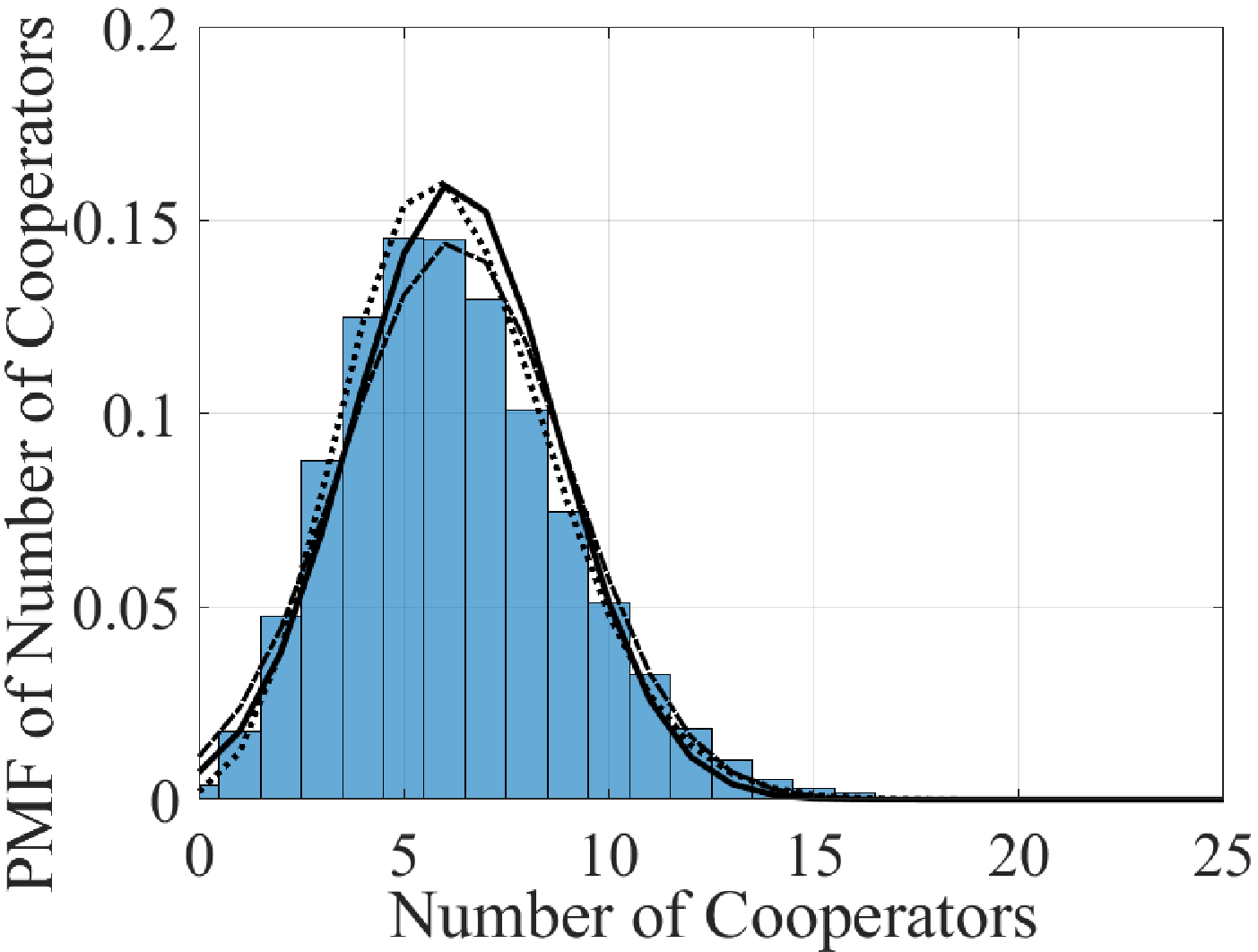}}
\subfigure[$R_1=150\,{\mu}\m$, $\eta=5$]{\label{fig4:6}\includegraphics[height=1.6in]{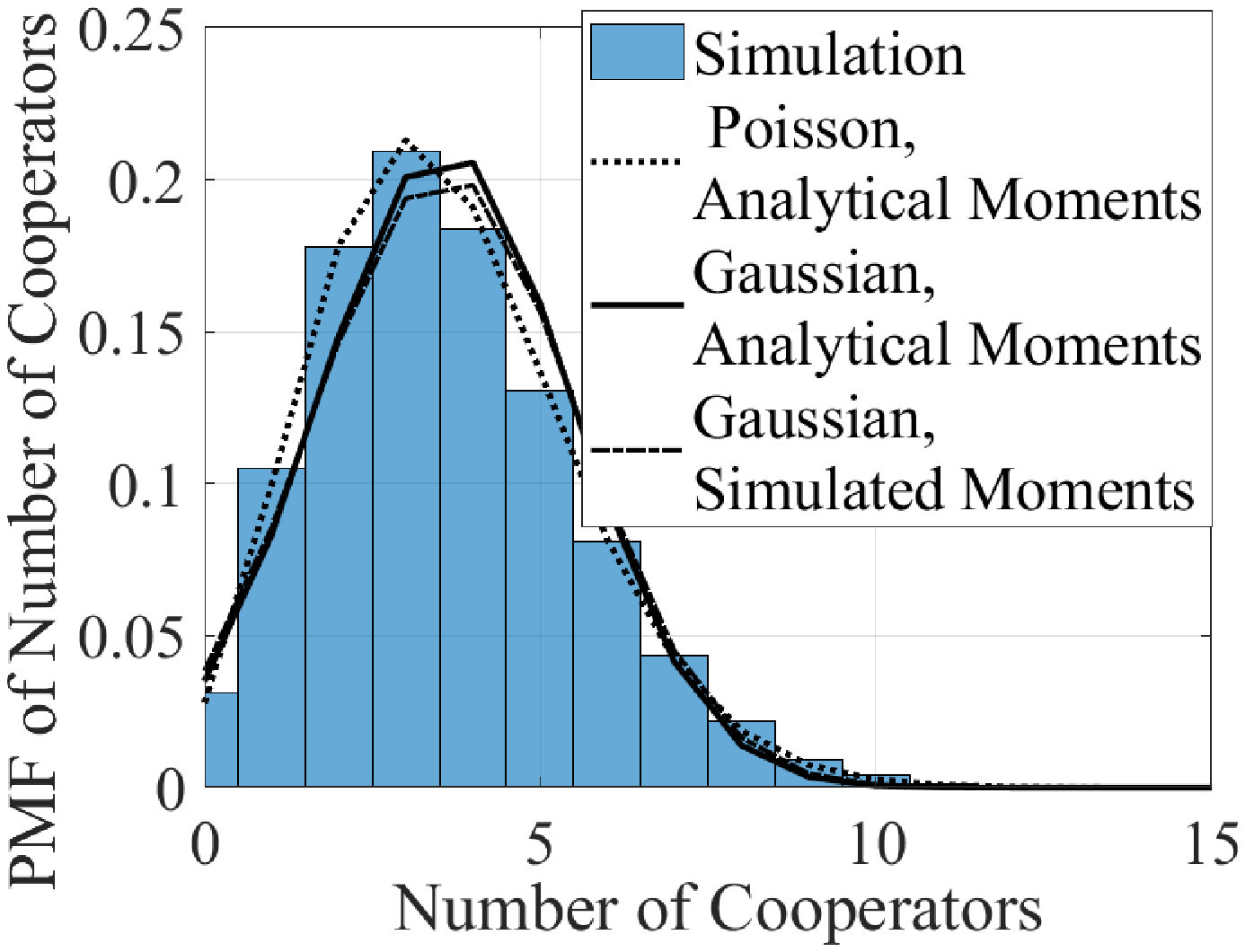}}
\vspace{-2mm}
\caption{The PMF of number of cooperators for different population radii $R_1$ and different thresholds $\eta$.}\label{fig4}
\vspace{-6mm}
\end{figure}

\begin{figure*}[!tbp]
		\centering
		\begin{minipage}[t]{0.49\textwidth}\hspace*{-5 mm}
			\centering
			\resizebox{1.05\linewidth}{!}{
				\includegraphics[scale=0.55]{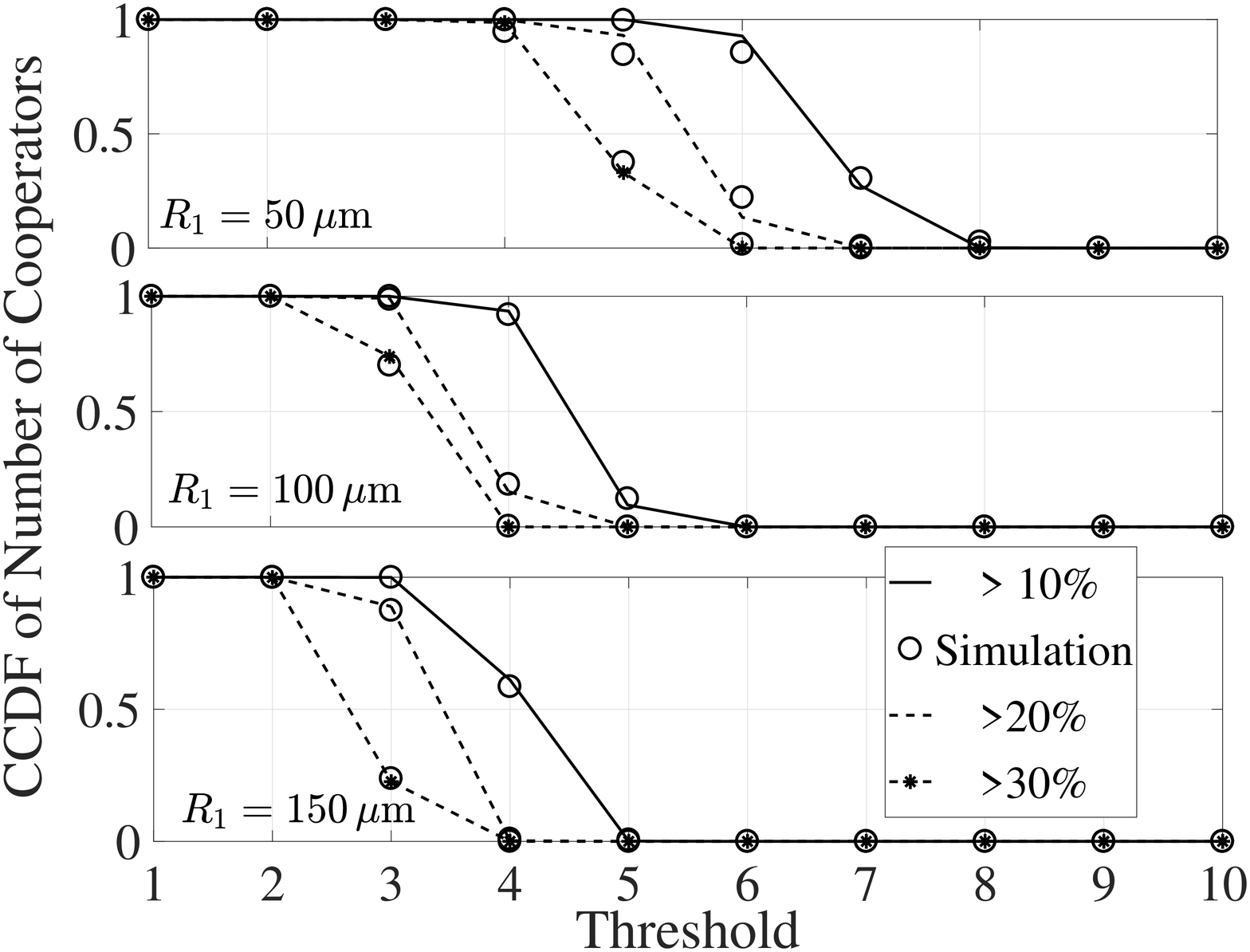}}\vspace*{-4 mm}
			\caption{
				The CCDF that the number of cooperators versus threshold $\eta$ for different population radii $R_1$.}
			\label{Fig-5}
		\end{minipage}
		\hfill
		\vspace*{-1 mm}
		\begin{minipage}[t]{0.49\textwidth}
			\centering
			\resizebox{1.05\linewidth}{!}{
				\includegraphics[scale=0.55]{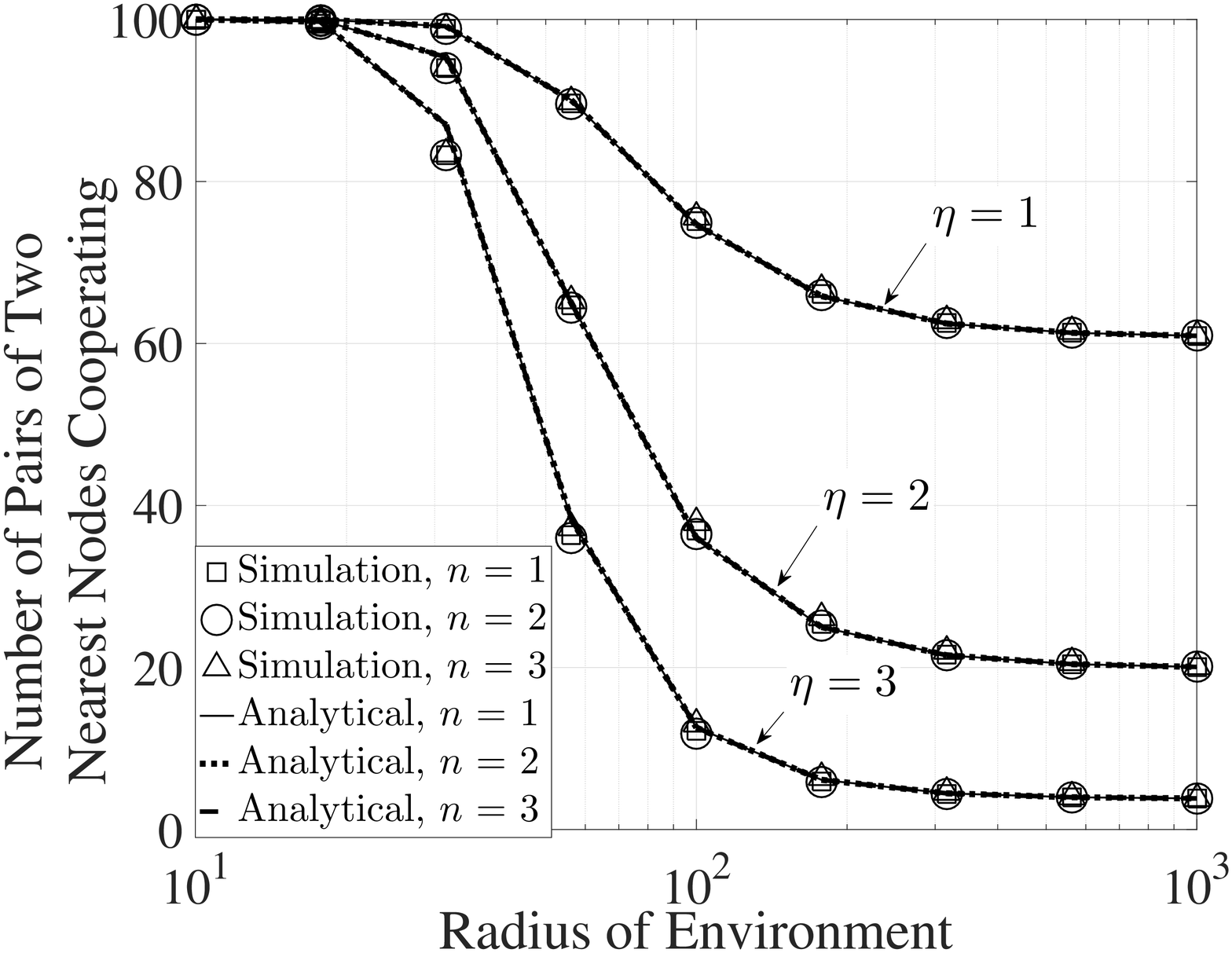}}\vspace*{-4 mm}
			\caption{
				The number of pairs of any node and its $n$th nearest node both cooperating versus the population radius. 
			}
			\label{Fig-6}
		\end{minipage}
		\vspace{-7 mm}
	\end{figure*}

The simulation details are as follows. The simulation environment is unbounded. We vary density, bacteria community radius $R_1$, and threshold $\eta$. Unless specified otherwise, we consider molecule degradation with rate $k=1\times10^{1}/{\s}$ in the environment, a circular RX with $R_{0}=0.757\,{\mu}\m$, emission rate $q=1 \times10^{3}\,{\mole}/{\s}$, and diffusion coefficient $D=5.5\times10^{-10}{\m^{2}}/{\s}$. 
The values of environment parameters are chosen to be on the same orders as those used in \cite{Danino2010,Trovato2014,Surette7046,Dilanji2012QuorumAA}. In particular, the chosen value of $D$ is the diffusion coefficient of the 3OC6-HSL in water at room temperature \cite{Dilanji2012QuorumAA}. The volume of a sphere with the chosen radius is approximately equal to the volume of \emph{V. fischeri}\footnote{In fact, the shape of \emph{V. fischeri} is a straight rod that is $0.8\,{\mu}\m$-$1.3\,{\mu}\m$ in diameter and $1.8\,{\mu}\m$-$2.4\,{\mu}\m$ in length. Due to its rod shape, we calculate its average volume as $\pi(((0.8\,{\mu}\m+1.3\,{\mu}\m)/2)/2)^2(1.8\,{\mu}\m+2.4\,{\mu}\m)/2=1.82\,{\mu}\m^3$. Thus, we choose $R_{0}=0.757\,{\mu}\m$ which satisfies  $(4/3)\pi{R_{0}}^3=1.82\,{\mu}\m^3$.}. We emphasize that these parameters are example values and  the general trends of our numerical observations in the following figures do not change for other combinations of parameter values. We simulate the Brownian motion of molecules using a particle-based method as described in \cite{Andrew2004}. The molecules are initialized at the center of bacteria. The location of each molecule is updated every time step $\Delta t$, where diffusion along each dimension is simulated by generating a normal RV with variance $2D\Delta t$. Every molecule has a chance of degrading in every time step with the probability $\exp(-k\Delta t)$. In simulations, the locations of bacteria are randomly generated according to a 2D PPP, thus the number of bacteria in different realizations is a Poisson RV and we consider that the average number of TXs or bacteria is 100 in Figs. \ref{fig-2}--\ref{Fig-6}. \Fr{Since the time between consecutive events in a 1D PPP is exponentially distributed, we generate i.i.d. exponential RVs for each bacterium to simulate the time between consecutive molecule releases. There are also other methods to generate random release times, e.g., we could generate Poisson RVs as the numbers of release times within a fixed time duration and then uniformly distribute these release times over the fixed duration.}
In Fig. \ref{fig-1}, there is one TX at a fixed location and for each realization we randomly generate molecule release times at the TX. In Figs. \ref{fig-2}--\ref{Fig-6}, for each realization we randomly generate both the locations and molecule release times for all TXs (bacteria).

In Fig. \ref{fig-1}, we plot the expected number of molecules observed at the RX due to one TX's impulse emission with $10^6$ molecules in Fig. \ref{fig-1b} and one TX's continuous emission in Fig. \ref{Fig-1a}. The analytical curves in Case i)--Case v) are obtained by \eqref{point-circle, impulse2}, \eqref{point-circle,self2}, \eqref{point1}, \eqref{point,no deg}, and \eqref{point-circle,self}, respectively. In Fig. \ref{fig-1b}, we see that there is an optimal time at which channel response is maximal when the RX is not at the TX, while the channel response always decreases with time when the RX is at the TX. This is not surprising since the molecules diffuse away once released. In Fig. \ref{Fig-1a}, we see that the channel response with molecular degradation converges as time goes to infinity, while the channel response without molecular degradation always increases with time. 

In Fig. \ref{fig-2}, we plot the expected number of molecules observed at the RX in Fig. \ref{fig-1c} and the corresponding cooperating probability at the RX in Fig. \ref{Fig-2a} due to continuous emission at randomly-distributed TXs for different environmental radii. We first discuss the results in Fig. \ref{fig-1c}. The asymptotic curves when the RX is at $(0,0)$ with the UCA and without UCA are obtained by \eqref{circle-cirle, agg4} and \eqref{circle-cirle, agg3}, respectively. The asymptotic curves when the RX is at $(\frac{R_1}{2},\frac{R_1}{2})$ with UCA and without UCA are obtained by \eqref{circle-cirle, agg2} and \eqref{circle-cirle, agg}, respectively. As observed in  Fig. \ref{Fig-1a}, we see that the expected number of molecules observed in Fig. \ref{fig-1c} first increases with time and then becomes stable after some time. We then see that the asymptotic curves with UCA and without UCA almost overlap with each other. This demonstrates the accuracy of the UCA in the derivation of the asymptotic channel response where a circular field of TXs continuously emits molecules.

Next, we discuss the results in Fig. \ref{Fig-2a}. The exact and approximate analytical curves are obtained by \eqref{prob,Laplace,fina} via \eqref{circle-cirle,indi2} and \eqref{prob,mean,poisson} via \eqref{circle-cirle, agg2}, respectively. We see that \eqref{prob,Laplace,fina} is always accurate while \eqref{prob,mean,poisson} is only accurate when the probability of cooperation is relatively high, e.g., $\widetilde{\prob}\left(N_{\agg}^{\dag}(\vec{x_i}|\lambda)\geq\eta\right)\geq10^{-1}$. We note that the computational complexity of \eqref{prob,mean,poisson} is much lower than that of \eqref{prob,Laplace,fina}. Thus, in the circumstances of limited computational capabilities and high probability of cooperation, \eqref{prob,mean,poisson} is a good method to estimate the probability of cooperation. Finally, we note that when $R_1$ decreases, the expected number of molecules and the probability of cooperation increase. This is because the density of TXs is higher when $R_1$ is smaller.



\Fr{In Fig. \ref{fig:spatial}, we simulate the decisions of bacteria under one realization of randomly-distributed bacteria locations and random molecule release times at all bacteria. We plot the spatial distribution of cooperators in this realization.} 
In Fig. \ref{Fig-2b}, we plot the first moment (i.e, the mean) of the number of cooperative bacteria versus threshold for different population radii. The exact analytical curves are obtained by \eqref{mean,fina} via \eqref{prob,Laplace,fina} and  \eqref{circle-cirle,indi2} and the approximate analytical curves are obtained by \eqref{mean,fina} via \eqref{prob,mean,poisson} and \eqref{circle-cirle, agg2}. We see that the curves obtained by \eqref{mean,fina} via \eqref{prob,mean,poisson} are only accurate when ${\E}_{\Phi}\{Z\}\geq10$. This is because \eqref{prob,mean,poisson} is only accurate when $\widetilde{\prob}\left(N_{\agg}^{\dag}(\vec{x_i}|\lambda)\geq\eta\right)\geq10^{-1}$, as observed in Fig. \ref{Fig-2a}. We also see that the analytical mean obtained by \eqref{mean,fina} via \eqref{prob,Laplace,fina} exactly matches with simulations. This observation numerically validates Remark \ref{remark:mean}, i.e., the approximation in \eqref{prob,app} is tight for the first moment of the number of cooperators. We also see that the expected number of cooperators decreases when the threshold increases, because the probability of cooperation is smaller when the threshold is higher, as observed in Fig. \ref{Fig-2a}.

In Figs. \ref{fig3:a}--\ref{fig3:c}, we plot the variance and moments of number of cooperators versus threshold for different population radii. The analytical variances are obtained by \eqref{var} and the analytical moments are obtained by \eqref{moment,relation,1} via \eqref{mean,fina}. We first see that when the population density is smaller (i.e., $R_1$ is larger), the accuracy of the analytical variances and the moments improves, thereby validating Remark~\ref{remark:mean}. We then see that the curves of different moments of number of cooperators merge as the threshold increases. This can be explained by the extreme case that the different moments of the number of cooperators would tend to zero as the threshold continually increases, leading to the fact that different moments of the number of cooperators become the same as the threshold continually increases.

\begin{table}[!t]
\renewcommand{\arraystretch}{1.1}
\centering
\caption{\Fr{Deviation between Simulation and Analytical Values at Peak PMF in Fig. \ref{fig4}}.}\label{tab:table1-ch4}
\vspace{-3mm}
\begin{tabular}{c||c|c|c|c|c|c}
\hline
  &  \bfseries(a) &  \bfseries(b)&  \bfseries(c)&  \bfseries(d)&  \bfseries(e)&  \bfseries(f)\\\hline\hline
Poisson Approximation & $6.01\%$ & $5.09\%$& $6.02\%$&$45.42\%$ & $9.65\%$& $1.77\%$ \\\hline
Gaussian Approximation & $6.01\%$ & $5.09\%$& $6.04\%$&$48.23\%$&$10.20\%$ &$4.01\%$\\\hline\end{tabular}
\vspace{-6mm}
\end{table}

\Fr{We investigate the distributions of the simulated number of cooperators using the statistical distribution fitting software EasyFit\footnote{ http://www.mathwave.com/easyfit-distribution-fitting.html}. We find that the simulated number of cooperators can be generally well fitted by Beta, Johnson SB, Normal (Gaussian), and Gamma distributions when the number of cooperators is large. When the number of cooperators is small, the simulated number of cooperators can be generally well fitted by Poisson and Binomial distributions. We note that Gaussian and Poisson distributions are the most convenient distributions among them while achieving good fitting accuracy. Hence, we use the Gaussian and Poisson distributions with our derived mean and variance to analytically fit the simulated distribution of cooperators in Fig. \ref{fig4} for assessing its accuracy to analytically predict the distribution of cooperators.}

In Fig. \ref{fig4}, we use the Poisson and Gaussian distributions with analytical mean ${\E}_{\Phi}\{Z\}$ and variance $\var\{Z\}$ shown in Fig. \ref{fig3} to fit the PMF of simulated number of cooperators. \Fr{To quantitatively assess the accuracy of Poisson and Gaussian approximations, we calculate the deviation between the simulation and analytical curves by ${|\textrm{analysis}-\textrm{simulation}|}/{\textrm{simulation}}$. We are interested in the deviation at the peak PMF since such deviation is the largest difference across the whole range of the PMF. The calculated deviation for different cases in Fig.~\ref{fig4} is listed in Table~\ref{tab:table1-ch4}.} Based on Table~\ref{tab:table1-ch4}, we see that the distribution of the number of cooperators can generally be well approximated by the Poisson and Gaussian distributions, especially when the expected number is relatively large, which meets our expectations discussed in Sec. \ref{subsec:mo}. When the number of cooperators is relatively small, e.g., $Z<15$, the Poisson approximation has better accuracy than the Gaussian approximation. This observation is also expected since the continuous Gaussian distribution is an approximation of the discrete distribution and such approximation is more accurate when the number of cooperators is higher. The deviation between the Poisson and Gaussian distributions and simulated distribution for $R_1=50\,{\mu}\m$ and $\eta=5$ is caused by the deviation between the analytical variance and simulated variance, as observed in Fig. \ref{fig3:a}.


In Fig. \ref{Fig-5}, we plot the complementary CDF (CCDF) of the number of cooperators versus threshold for different population radii. \Fr{These CCDFs could be interpreted as the probability of success to activate a colony of bacteria if we can define the activation of the colony as a minimal number of cells (i.e., fraction of the colony) being activated. The probability of success indicates the overall activation level of a colony, e.g., the likelihood of the colony forming a biofilm.} The analytical curves are obtained by the CCDF of the Poisson distribution with analytical mean ${\E}_{\Phi}\{Z\}$ and variance $\var\{Z\}$ shown in Fig. \ref{fig3}, respectively. We see that the CCDF of the number of cooperators can be well approximated by that of the Poisson distributions. We also see that the CCDF of the number of cooperators decreases as the threshold increases. The PMF and CDF in Figs. \ref{fig4} and \ref{Fig-5} indicates that the Poisson and Gaussian approximations can be used to not only accurately model the distribution of the number of observed molecules as many MC works have done, e.g., \cite{6712164,Lin2015,8742793}, but also model the distribution of the number of cooperative bacteria in QS with good accuracy. Therefore, the easy-to-use Poisson and Gaussian distributions with our derived mean and variance serve as powerful tools to predict the group behavior of bacteria in QS and control their group behavior by adjusting the environmental parameters, e.g., diffusion coefficient and chemical reaction rate. Thus, our research plays as an important role in advancing numerous QS-related healthcare and environmental applications, e.g., preventing the formation of biofilms in antibiotic resistance and understanding the bioluminescene in environmental monitoring.


In Fig. \ref{Fig-6}, we plot the number of pairs of any node and its $n$th nearest node both cooperating $P(n)$ versus the population radius $R_1$ for different thresholds $\eta$. The analytical curves are obtained by \eqref{near3}. We first see that for the same threshold $\eta$, the curves of $P(n)$ with different $n$ almost overlap. This is because bacteria are randomly distributed and the observations among different bacteria are independent. This observation is not intuitive, which suggests that the distance between bacteria has a minor impact on the average number of clusters of cooperators. Second, we see that the curves of $P(n)$ first decrease and then converge to a constant number as the population radius $R_1$ increases. This is because when the population radius increases, the number of molecules observed by the bacteria decreases, but as the population radius tends to infinity, the molecules received by any bacterium is dominated by the molecules released from itself and the number of molecules received by any bacterium converges to a constant number. This observation suggests that the density of bacteria has a marginal impact on the average number of clusters of cooperators when the density is very low. 



\vspace{-4mm}
\section{Conclusions}\label{sec:con}

In this work, we provided an analytically-tractable model for predicting the concentration of molecules observed by bacteria and the statistics of the number of responsive cooperative bacteria, by taking the motion of molecules undergoing independent diffusion and degradation into consideration. We adopted some assumptions to capture the basic features of QS, e.g., random bacterial location, random molecular emission times, and each bacterium both acting as a TX and a RX. Upon these realistic assumptions, the 2D channel response and the expected probability of cooperation at a bacterium due to continuous emission of molecules at randomly-distributed bacteria were derived. The different order moments and cumulants, the CDF and PMF of the number of cooperators, and the number of pairs of two bacteria both cooperating, were also derived. Since we considered molecular propagation channels among a population of bacteria, the impact of environmental factors on the QS process could be predicted, e.g., environment temperature and pH can affect the diffusion coefficient and degradation rate, respectively.

Interesting future works include relaxing these simplifications, and validating our theoretical model using lab experiments in a 2D environment. \Fr{Potential experimental setups for validation may include bacteria strains in plastic Petri dishes, bioluminescence monitoring systems that detect cooperative bacteria concentrations by their light emission \cite{Cloak4666}, and computers for post-processing and analysis.
Modeling multiple species of bacteria in QS is also interesting future work since multispecies QS signals may inhibit quorum sensing activation [2]. For example, some types of bacteria produce enzymes that destroy quorum sensing signals of other types of bacteria. Such inhibition could lead to lower cooperative probabilities and fewer cooperators compared to the case when QS is in the absence of the bacteria that release the destructive enzymes.} Finally, we note that our analytically-tractable model could be readily extended to a three-dimensional (3D) environment by changing the 2D area integrations to 3D volume integrations.




%
%
\numberwithin{equation}{section}
\begin{appendices}
\vspace{-6mm}
\section{Proof of \proref{propo:impulse-to-cont}}\label{Proof of impulse to conti}
\Fr{To prove \eqref{cont-impulse}, we have
\begin{align}\label{eq:Re1-2}
\overline{N}\left(\vec{b},t\right)\overset{(a)}{=}\E\Bigg\{\sum_{\tau\in [0,t]}\overline{N}_{\im}\left(\vec{b},t-\tau\right)\Bigg\}\overset{(b)}{=}q\int_{\tau=0}^{t}\overline{N}_{\im}\left(\vec{b},t-\tau\right)d\tau
\overset{(c)}{=}q\int_{\tau=0}^{t}\overline{N}_{\im}\left(\vec{b},\tau\right)d\tau,
\end{align}
where $\overline{N}_{\im}\left(\vec{b},t-\tau\right)$ is the channel response observed at time $t$ due to the one molecule emitted at time $\tau$ and $\tau$ is a random time instant on a real line $[0,t]$ distributed according to a 1D PPP with density (rate) $q$. Equality (a) exploits the fact that the channel response at time $t$ due to \emph{continuous} emission is equal to the expected sum of all channel responses due to all impulse emissions at different random time instants. Equality (b) is obtained by applying Campbell's Theorem and equality (c) is obtained by applying a variable transformation. By taking the limit $t\rightarrow\infty$ in \eqref{eq:Re1-2}, we obtain \eqref{cont-impulse}.} 




\vspace{-6mm}
\section{Proof of \thmref{Theorem:impulse,any} and \thmref{Theorem:impulse,0}}\label{Chap4:Proof of impulse}
To prove \eqref{point-circle, impulse2}, 
we have

\begin{align}\label{point-circle, impulse}
\overline{N}_{\im}\left(\vec{b},\tau\right) \overset{(a)}{=}&\; \int_{r=0}^{R_0}\int_{\theta=0}^{2\pi}C\left(\vec{r_1},\tau\right)rd\theta dr
=\int_{r=0}^{R_0}\int_{\theta=0}^{2\pi}\frac{1}{(4\pi D \tau)}\exp\left(-\frac{|\vec{r_1}|^{2}}{4D\tau}-k\tau\right)rd\theta dr\nonumber\\
=&\;\int_{r=0}^{R_0}\int_{\theta=0}^{2\pi}\frac{1}{(4\pi D \tau)}\exp\left(-\frac{{|\vec{b}|}^2+r^2+2{|\vec{b}|}r\cos\theta}{4D\tau}-k\tau\right)rd\theta dr\nonumber\\
\overset{(b)}{=}&\;\frac{1}{4\pi D}\exp(-k\tau)\int_{r=0}^{R_0}\frac{r}{\tau}\exp\left(-\frac{{|\vec{b}|}^2+r^2}{4D\tau}\right)2\pi I_0\left(\frac{|\vec{b}| r}{2D\tau}\right)dr\nonumber\\
\overset{(c)}{\approx} &\;\sum_{i=1}^{4}\bigg\{\frac{1}{2D}\exp(-k\tau)\int_{r=0}^{R_0}\frac{r}{\tau}\exp\bigg(-\frac{{|\vec{b}|}^2+r^2}{4D\tau}\bigg)\alpha_i\exp\left(\beta_i \frac{|\vec{b}|r}{2D\tau}\right)dr\bigg\}
\end{align}
where $C\left(\vec{r},\tau\right) = \frac{1}{\left(4\pi D \tau\right)}\exp\left(-\frac{|\vec{r}|^{2}}{4D\tau}-k\tau\right)$ \cite[eq.~(3.4)]{Crank_book} is the channel response at the point defined by $\vec{r}$ at the time $\tau$ due to an impulse emission of one molecule from the point at $(0,0)$ at time $\tau=0$ into an unbounded 2D environment. 
Equality (a) is due to the fact that $\overline{N}_{\im}\left(\vec{b},\tau\right)$ for a circular passive observer $S_0$ centered at $\vec{b}$ can be obtained by integrating $C\left(\vec{r_1},\tau\right)$ over $S_0$,
where $\vec{r_1}$ is a vector from $(0,0)$ to a point within the RX $S_0$. Equality (b) is obtained by applying \cite[eq.~(3.339)]{gradshteyn2007} and
equality (c) is obtained by applying $I_0(z)\approx\sum_{i=1}^{4}\alpha_i\exp(\beta_i z)$ \cite[eq.~(7)]{Salahat2013}. Calculating the final expression of \eqref{point-circle, impulse}, we obtain \eqref{point-circle, impulse2}.
To prove \eqref{point-circle,self2}, we simplify \eqref{point-circle, impulse} using $|\vec{b}|=0$ as
\begin{align}\label{point-circle,self1}
\overline{N}_{\im,\self}\left(\tau\right)
=\int_{r=0}^{R_0}\int_{\theta=0}^{2\pi}\frac{\exp\left(-\frac{r^2}{4D\tau}-k\tau\right)}{(4\pi D \tau)}\,d\theta \,dr
=\int_{r=0}^{R_0}\frac{r}{(2D \tau)}\exp\left(-\frac{r^2}{4D\tau}-k\tau\right) \,dr.
\end{align}

We then apply \cite[eq.~(2.33.12)]{gradshteyn2007}
to \eqref{point-circle,self1} to solve it as \eqref{point-circle,self2}, which completes the proof.

\vspace{-5mm}
\section{Proof of \thmref{Theorem:continuous,any} and \thmref{Theorem:continuous,0}}\label{Proof:Cont-any}
\Fr{We first prove \eqref{point1}. Based on \eqref{cont-impulse}, the channel response due to continuous emission in \eqref{point1} can be obtained by integrating the impulse channel response over time,
but integrating \eqref{point-circle, impulse2} over $\tau$ incurs very high complexity. Thus, we simplify \eqref{point-circle, impulse2} by considering UCA within the circular RX. i.e., $\overline{N}_{\im}\left(\vec{b},\tau\right) \approx \pi R_0^2C\left(\vec{b},\tau\right)$. 
Based on \eqref{cont-impulse} and the UCA, we evaluate $\overline{N}\left(\vec{b},\infty\right)$ as
\begin{align}\label{point}
\overline{N}\left(\vec{b},\infty\right) \approx \pi R_0^2\int_{\tau=0}^{\infty}qC\left(\vec{b},\tau\right)d\tau
\approx \pi R_0^2 q \int_{\tau=0}^{\infty}\frac{1}{\left(4\pi D \tau\right)}\exp\left(-\frac{|\vec{b}|^{2}}{4D\tau}-k\tau\right)d\tau.
\end{align}}

\Fr{We then employ
$\int_{0}^{\infty} x^{\nu-1}\exp\left(-\frac{\beta}{x}-\gamma x\right)dx = 2\left(\frac{\beta}{\gamma}\right)^{\frac{\nu}{2}}K_{\nu}(2\sqrt{\beta\gamma})$
\cite[eq.~(3.471)]{gradshteyn2007} to solve \eqref{point} as \eqref{point1}.} To prove \eqref{point-circle,self}, we apply \eqref{point-circle,self2}, $\int_{=0}^{\infty}\exp(-px)dx = {1}/{p}$ \cite[eq.~(3.310)]{gradshteyn2007}, and $\int_{0}^{\infty} \exp\left(-\frac{\beta}{x}-\gamma x\right)dx = \frac{\beta}{\gamma}K_{1}(\sqrt{\beta\gamma})$ \cite[eq.~(3.324.1)]{gradshteyn2007} 
to \eqref{cont-impulse}, we evaluate $\overline{N}_{\self}\left(\infty\right)$ as \eqref{point-circle,self}. This completes the proof.

\vspace{-5mm}
\section{Proof of \thmref{Theorem:aggregate_mole}}\label{Chap4:aggregate_mole}
Using Campbell's theorem \cite{Haenggi:2012:SGW:2480878}, we first write
\begin{align}\label{circle-cirle, agg,p2}
{\E}_{\Phi}\bigg\{\overline{N}_{\agg}\left(\vec{b}|\lambda\right)\bigg\}
={\E}_{\Phi}\Bigg\{\sum_{\vec{a}\in\Phi(\lambda)}\overline{N}\Big(\vec{b}|\vec{a}\Big)\Bigg\}
=\int_{|\vec{r}|=0}^{R_1}\int_{\varphi=0}^{2\pi}\overline{N}\Big(\vec{b}|\vec{r}\Big)\lambda |\vec{r}|\,d\varphi\,d|\vec{r}|,
\end{align}
where $\vec{r}$ is a vector from $(0,0)$ to a point within the population circle $S_1$ and $\varphi$ is the supplement of the angle between $\vec{r}$ and $\vec{b}$. 
We note that $\overline{N}\left(\vec{b}|\vec{r}\right)$ is obtained by multiplying the point channel response by the emission rate $q$, integrating over $S_0$, and then integrating over all time, i.e.,
\begin{align}\label{circle-cirle,indi}
\overline{N}\left(\vec{b}|\vec{r}\right) = &\int_{\tau=0}^{\infty}\int_{|\vec{r_0}|=0}^{R_0}\int_{\theta=0}^{2\pi}qC\left(\vec{d},\tau\right)|\vec{r_0}|\,d\theta\,d|\vec{r_0}|\,d\tau,\nonumber\\
=&\int_{\tau=0}^{\infty}\int_{|\vec{r_0}|=0}^{R_0}\int_{\theta=0}^{2\pi}\frac{q}{(4\pi D \tau)}\exp\left(-\frac{|\vec{d}|^{2}}{4D\tau}-k\tau\right)|\vec{r_0}|\,d\theta\,d|\vec{r_0}|\,d\tau,\nonumber\\
=&\int_{\tau=0}^{\infty}\!\int_{|\vec{r_0}|=0}^{R_0}\!\int_{\theta=0}^{2\pi}\!\frac{q}{(4\pi D \tau)}\exp\left(\!-\frac{|\vec{l}|^2+|\vec{r_0}|^2+2|\vec{l}||\vec{r_0}|\cos\theta}{4D\tau}-k\tau\!\right)|\vec{r_0}|\,d\theta\,d|\vec{r_0}|\,d\tau,
\end{align}
where $\vec{l}$ is a vector from $\vec{r}$ to $\vec{b}$, i.e., $\vec{l}=\vec{b}-\vec{r}$, $\vec{r_0}$ is a vector from $\vec{b}$ to a point within the RX circle $S_0$, $\vec{d}$ is a vector from $\vec{r}$ to $\vec{r_0}$, and $\theta$ is the supplement of the angle between $\vec{l}$ and $\vec{r_0}$. According to the law of cosines, we obtain $|\vec{l}|^{2}={|\vec{b}|}^2+|\vec{r}|^2+2{|\vec{b}|}|\vec{r}|\cos\varphi$ and $|\vec{d}|^{2}=|\vec{l}|^{2}+|\vec{r_0}|^2+2|\vec{l}||\vec{r_0}|\cos\theta$. We then apply $|\vec{l}|=\sqrt{{|\vec{b}|}^2+|\vec{r}|^2+2{|\vec{b}|}|\vec{r}|\cos\varphi}$ to rewrite \eqref{circle-cirle,indi} as
\begin{align}\label{circle-cirle,indi1}
\overline{N}\left(\vec{b}|\vec{r}\right) = &\;\int_{\tau=0}^{\infty}\int_{|\vec{r_0}|=0}^{R_0}\int_{\theta=0}^{2\pi}\frac{q}{(4\pi D \tau)}\exp\left(-\frac{\Upsilon^2(\vec{b})}{4D\tau}-k\tau\right)|\vec{r_0}|\,d\theta\,d|\vec{r_0}|\,d\tau\nonumber\\
\overset{(a)}{=} &\;\int_{|\vec{r_0}|=0}^{R_0}\int_{\theta=0}^{2\pi}\frac{q}{2D\pi}K_0\left(\sqrt{\frac{k}{D}} \Upsilon(\vec{b})\right)|\vec{r_0}|\,d\theta\,d|\vec{r_0}|,
\end{align}
where equality (c) is obtained by applying \cite[eq.~(3.471)]{gradshteyn2007}.
We finally substitute \eqref{circle-cirle,indi1} into \eqref{circle-cirle, agg,p2} and arrive at \eqref{circle-cirle, agg}. This completes the proof.

\vspace{-5mm}
\section{Proof of Corollary \ref{corollary:aggregate_mole-UCA}, Corollary \ref{corollary:aggregate_mole-center}, and Corollary \ref{corollary:aggregate_mole-center-UCA}}\label{proof:aggregate_mole-UCA}
We first prove \eqref{circle-cirle, agg2}. Using UCA, we have
\begin{align}\label{circle-cirle,indi2}
\overline{N}\left(\vec{b}|\vec{r}\right) \approx&\; \left(\int_{\tau=0}^{\infty}qC\left(\vec{l},\tau\right)\,d\tau\right)\pi {R_0}^2
\approx\int_{\tau=0}^{\infty}\frac{q \pi{R_0}^2}{(4 \pi D \tau)}\exp\left(-\frac{{|\vec{b}|}^2+|\vec{r}|^2+2{|\vec{b}|}|\vec{r}|\cos\varphi}{4D\tau}-k\tau\right)d\tau,\nonumber\\
\approx &\;\frac{q{R_0}^2}{2D}K_0\left(\sqrt{\frac{k}{D}\Omega(\vec{b})}\right).
\end{align}

We then substitute \eqref{circle-cirle,indi2} into \eqref{circle-cirle, agg,p2} and obtain
\eqref{circle-cirle, agg2}. \eqref{circle-cirle, agg3} can be proven by applying $|\vec{b}|=0$ to \eqref{circle-cirle, agg}.
To prove \eqref{circle-cirle, agg4}, 
we apply $|\vec{b}|=0$ to \eqref{circle-cirle, agg2} to rewrite \eqref{circle-cirle, agg2} as \eqref{circle-cirle, agg4-1}. By evaluating the final expression of \eqref{circle-cirle, agg4-1}, we obtain \eqref{circle-cirle, agg4}.
\begin{align}\label{circle-cirle, agg4-1}
{\E}_{\Phi}\{\overline{N}_{\agg}(\vec{b}|\lambda)\}\Big{|}_{|\vec{b}|=0} \!\!\approx \!\! \int_{|\vec{r}|=0}^{R_1}\!\int_{\varphi=0}^{2\pi}\!\frac{q{R_0}^2}{2D}K_0\!\left(\!\!\sqrt{\frac{k}{D}} |\vec{r}|\!\right)\!\lambda |\vec{r}|\,d\varphi\,d|\vec{r}|
\!\approx\! \int_{|\vec{r}|=0}^{R_1}\!\!\frac{q\pi{R_0}^2}{D}K_0\!\left(\!\!\sqrt{\frac{k}{D}} |\vec{r}|\!\right)\!\lambda |\vec{r}|\,d|\vec{r}|.
\end{align}

\vspace{-6mm}
\section{Proof of Lemma \ref{lemma:coop-prob}}\label{proof:coop-prob}

We rewrite \eqref{prob,Laplace} as
\begin{align}\label{prob,Laplace1}
&\;\widetilde{\prob}\left(N_{\agg}^{\dag}(\vec{x_i}|\lambda)\geq\eta\right)
\overset{(a)}{=} 1-{\E}_{\Phi}\Bigg\{\sum_{n=0}^{\eta-1}\frac{1}{n!}\exp\left\{-\overline{N}_{\agg}^{\dag}(\vec{x_i}|\lambda)\right\}\left(\overline{N}_{\agg}^{\dag}(\vec{x_i}|\lambda)\right)^{n}\Bigg\}\nonumber\\
\!\overset{(b)}{=}\!
&\;1-\sum_{n=0}^{\eta-1}\frac{1}{n!}{\E}_{\Phi}\Bigg\{\frac{\partial^n\exp\left\{\overline{N}_{\agg}^{\dag}(\vec{x_i}|\lambda)\rho\right\}}{\partial \rho^n}\Bigg|_{\rho=-1}\Bigg\}
\!\overset{(c)}{=} \!1-\sum_{n=0}^{\eta-1}\frac{1}{n!}\frac{\partial^n {\E}_{\Phi}\Big\{\exp\left\{\overline{N}_{\agg}^{\dag}(\vec{x_i}|\lambda)\rho\right\}\Big\}}{\partial \rho^n}\Bigg|_{\rho=-1},
\end{align}
where equality (a) is due to the CDF of Poisson RV $N_{\agg}^{\dag}(\vec{x_i}|\lambda)$. Equality (b) is due to exchanging the order of sum and expectation and $\exp\left\{-\overline{N}_{\agg}^{\dag}(\vec{x_i}|\lambda)\right\}\left(\overline{N}_{\agg}^{\dag}(\vec{x_i}|\lambda)\right)^{n}=\frac{\partial^n\exp\left\{\overline{N}_{\agg}^{\dag}(\vec{x_i}|\lambda)\rho\right\}}{\partial \rho^n}\Bigg|_{\rho=-1}$ in \cite{8030318}. Equality (c) is due to exchanging the order of derivative and expectation.
Applying $\mathcal{L}_{\overline{N}_{\agg}^{\dag}(\vec{x_i}|\lambda)}(s)
= {\E}_{\Phi}\Big\{\exp\left\{-s\overline{N}_{\agg}^{\dag}(\vec{x_i}|\lambda)\right\}\Big\}$ to \eqref{prob,Laplace1}, we obtain \eqref{prob,Laplace3}.

\vspace{-6mm}
\section{Proof of Lemma \ref{Theorem:Laplace}}\label{Chap4:Proof of Laplace}
We first recall that the bacterium $i$ observes molecules in the environment released from all bacteria (also including the molecules released from itself). Thus, we have
\begin{align}\label{obsExp}
\overline{N}_{\agg}^{\dag}\left(\vec{x_i}|\lambda\right)=\sum_{\vec{x_j}\in\Phi\left(\lambda\right)}\overline{N}\left(\vec{x_i}|\vec{x_j}\right)
=\overline{N}\left(\vec{x_i}|\vec{x_i}\right)+\sum_{\vec{x_j}\in\Phi\left(\lambda\right)/\vec{x_i}}\overline{N}\left(\vec{x_i}|\vec{x_j}\right)
=\overline{N}_{\self}+\sum_{\vec{a}\in\Phi\left(\acute{\lambda}\right)}\overline{N}\left(\vec{x_i}|\vec{a}\right),
\end{align}
where $\overline{N}_{\self}$ is given in \eqref{point-circle,self}.
We consider a new density $\acute{\lambda}$ to keep the average number of bacteria the same after the approximation of \eqref{obsExp}. We then apply \eqref{obsExp} to $\mathcal{L}_{\overline{N}_{\agg}^{\dag}(\vec{x_i}|\lambda)}(s)$ to rewrite it as
\begin{align}\label{Laplace}
\mathcal{L}_{\overline{N}_{\agg}^{\dag}(\vec{x_i}|\lambda)}(s)
=&\;{\E}_{\Phi}\Bigg\{\exp\Bigg\{-s \Bigg\{\sum_{\vec{a}\in\Phi(\acute{\lambda})}\overline{N}(\vec{x_i}|\vec{a})+\overline{N}_{\self}\Bigg\} \Bigg\}\Bigg\},\nonumber\\
= &\;\exp\left(-s\overline{N}_{\self}\right){\E}_{\Phi}\Bigg\{\prod_{\vec{a}\in\Phi(\acute{\lambda})}\exp\left\{-s \overline{N}(\vec{x_i}|\vec{a})\right\}\Bigg\}.
\end{align}

Using the probability generating functional (PGFL) for the PPP \cite[eq.~(4.8)]{Haenggi:2012:SGW:2480878}, we rewrite \eqref{Laplace} as \eqref{Laplace1}. This completes the proof.

\vspace{-6mm}
\section{Proof of Theorem \ref{theorem:MGF-exact}}\label{proof:MGF-exact}
Using the definition of MGF \cite{ROSS201489}, the MGF of $Z$ is given by
${M}_{Z}(u)={\E}\{\exp(uZ)\}$.
We substitute $Z=\sum_{\vec{x_i}\in\Phi(\lambda)}B(\vec{x_i},\Phi)$ into ${M}_{Z}(u)={\E}\{\exp(uZ)\}$ to rewrite ${M}_{Z}(u)$ as
\begin{align}\label{mgf1}
{M}_{Z}(u)={\E}\Bigg\{\exp\Bigg(u\sum_{\vec{x_i}\in\Phi(\lambda)}B(\vec{x_i},\Phi)\Bigg)\Bigg\}
={\E}\Bigg\{\prod_{\vec{x_i}\in\Phi(\lambda)}\exp\Big(uB(\vec{x_i},\Phi)\Big)\Bigg\}.
\end{align}

\Fr{The expectation in \eqref{mgf1} is averaged over many realizations of randomly-distributed bacteria locations and their binary decisions. Thus, the expectation in \eqref{mgf1} can be written as first averaging over the binary decisions of bacteria and then averaging over the spatial point process $\Phi$.} By doing so, we rewrite \eqref{mgf1} as
\begin{align}\label{mgf2}
{M}_{Z}(u)=&\;{\E}_{\Phi}\Bigg\{\prod_{\vec{x_i}\in\Phi(\lambda)}{\E}_{B}\{\exp(uB(\vec{x_i},\Phi))\}\Bigg\}\nonumber\\
\overset{(a)}{=} &\;{\E}_{\Phi}\Bigg\{\prod_{\vec{x_i}\in\Phi(\lambda)}\{\exp(u)\prob\left(B(\vec{x_i},\Phi)=1\right)+\left(1-\prob\left(B(\vec{x_i},\Phi)=1\right)\right)\}\Bigg\}\nonumber\\
=&\;{\E}_{\Phi}\Bigg\{\prod_{\vec{x_i}\in\Phi(\lambda)}\{1+(\exp(u)-1)\prob\left(B(\vec{x_i},\Phi)=1\right)\}\Bigg\},
\end{align}
where equality (a) is because $B(\vec{x_i},\Phi)$ is a Bernoulli RV with mean $\prob\left(B(\vec{x_i},\Phi)=1\right)$. We recall that the bacterium $i$ is a cooperator, i.e., $B(\vec{x_i},\Phi)=1$, if $N_{\agg}^{\dag}\left(\vec{x_i}|\lambda\right)$ is larger than $\eta$. Thus, we derive $\prob\left(B(\vec{x_i},\Phi)=1\right)$ as
\begin{align}\label{prob}
\prob\left(B(\vec{x_i},\Phi)=1\right)= &\;\prob\left(N_{\agg}^{\dag}(\vec{x_i}|\lambda)\geq\eta|\overline{N}_{\agg}^{\dag}\left(\vec{x_i}|\lambda\right)\right),
\end{align}
where $\prob\left(N_{\agg}^{\dag}(\vec{x_i}|\lambda)\geq\eta|\overline{N}_{\agg}^{\dag}\left(\vec{x_i}|\lambda\right)\right)$ is the conditional cooperating probability for the bacterium $i$ in a given realization of the spatial random point process $\Phi$. Analogously to Sec.~\ref{subsec:exact probability}, we assume that $N_{\agg}^{\dag}(\vec{x_i}|\lambda)$ is a Poisson RV and apply $\overline{N}_{\agg}^{\dag}\left(\vec{x_i}|\lambda\right)=\sum_{\vec{x_j}\in\Phi\left(\lambda\right)}\overline{N}\left(\vec{x_i}|\vec{x_j}\right)$ to rewrite \eqref{prob} as
\begin{align}\label{prob1}
\prob\left(B(\vec{x_i},\Phi)=1\right)= &\;1-\left(\sum_{n=0}^{\eta-1}\frac{1}{n!}\exp\left\{-\overline{N}_{\agg}^{\dag}(\vec{x_i}|\lambda)\right\}\left(\overline{N}_{\agg}^{\dag}(\vec{x_i}|\lambda)\right)^{n}\right)\nonumber\\
=&\;1-\left(\sum_{n=0}^{\eta-1}\frac{1}{n!}\exp\Bigg\{-\sum_{\vec{x_j}\in\Phi(\lambda)}\overline{N}(\vec{x_i}|\vec{x_j})\Bigg\}\left(\sum_{\vec{x_j}\in\Phi(\lambda)}\overline{N}(\vec{x_i}|\vec{x_j})\right)^{n}\right).
\end{align}

We finally substitute \eqref{prob1} into \eqref{mgf2}, we obtain \eqref{mgf3}.
\vspace{-4mm}
\section{Proof of Theorem \ref{theorem:MGF-approximate}}\label{proof:approximate}
To obtain \eqref{mgf, app1}, we use the expected cooperating probability over the spatial point process $\Phi$ to approximate the conditional cooperating probability for a given instantaneous realization of this point process $\Phi$. By doing so, we approximate \eqref{prob} as
\begin{align}\label{prob,app}
\prob\left(B(\vec{x_i},\Phi)=1\right)\approx &\;{\E}_{\Phi}\Big\{\prob\left(B(\vec{x_i},\Phi)=1\right)\Big\},
\end{align}
\begin{align}\label{prob,app1}
{\E}_{\Phi}\Big\{\prob\left(B(\vec{x_i},\Phi)=1\right)\Big\}= &\;{\E}_{\Phi}\Big\{\prob\left(N_{\agg}^{\dag}(\vec{x_i}|\lambda)\geq\eta|\overline{N}_{\agg}^{\dag}(\vec{x_i}|\lambda)\right)\Big\}
= \widetilde{\prob}\left(N_{\agg}^{\dag}(\vec{x_i}|\lambda)\geq\eta\right),
\end{align}
where $\widetilde{\prob}\left(N_{\agg}^{\dag}(\vec{x_i}|\lambda)\geq\eta\right)$ is evaluated in Sec. \ref{PrCoop}. The approximated $\prob\left(B(\vec{x_i},\Phi)=1\right)$ in \eqref{prob,app} only depends on the location $\vec{x_i}$ and does not depend on the position of other bacteria in $\Phi$.

We then substitute \eqref{prob,app} into \eqref{mgf2} and obtain the approximated ${M}_{Z}(u)$ as
\begin{align}\label{mgf, app}
{M}_{Z}(u)
\approx&\;{\E}_{\Phi}\Bigg\{\prod_{\vec{x_i}\in\Phi(\lambda)}\{1+(\exp(u)-1)\widetilde{\prob}\left(N_{\agg}^{\dag}(\vec{x_i}|\lambda)\geq\eta\right)\}\Bigg\}.
\end{align}

Using PGFL \cite[eq.~(4.8)]{Haenggi:2012:SGW:2480878} for a PPP, we derive \eqref{mgf, app} as \eqref{mgf, app1}. Based on \eqref{mgf, app1}, we obtain \eqref{cgf} via $\mathcal{K}_{Z}(u)=\log{\E}_{\Phi}\{\exp(uZ)\}$. This completes the proof.


%
\vspace{-4mm}
\section{Proof of Proposition \ref{mean1}}\label{proof:mean1}
Recalling $Z=\sum_{\vec{x_i}\in\Phi(\lambda)}B(\vec{x_i},\Phi)$, we directly write ${\E}_{\Phi}\{{Z}\}$ (instead of using the MGF of $Z$) as
\begin{align}\label{coop,de}
{\E}_{\Phi}\{{Z}\}={\E}_{\Phi}\{\overline{Z}\}= &\;{\E}_{\Phi}\Bigg\{\sum_{\vec{x_i}\in\Phi(\lambda)}\prob\left(B(\vec{x_i},\Phi)=1\right)\Bigg\},
\end{align}
where $\overline{Z}$ is the mean of $Z$ for a given instantaneous realization of $\Phi$. Applying Campbell-Mecke's theorem of PPPs \cite[eq.~(8.7)]{Haenggi:2012:SGW:2480878} given by
${\E}_{\Phi}\Big\{\sum_{x\in\Phi} h(x,\Phi)\Big\}= \lambda\int_{\mathbb{R}^2}{{\E}_{\Phi}}(h(x,\Phi))dx$
to \eqref{coop,de}, we rewrite \eqref{coop,de} as
\begin{align}\label{coop}
&\;{\E}_{\Phi}\Bigg\{\sum_{\vec{x_i}\in\Phi(\lambda)}\prob\left(B(\vec{x_i},\Phi)=1\right)\Bigg\}
= \int_{|\vec{r_1}|=0}^{R_1}{{\E}_{\Phi}}\{\prob\left(B(\vec{r_1},\Phi)=1\right)\}\lambda 2\pi|\vec{r_1}|d|\vec{r_1}|.
\end{align}
Applying \eqref{prob} and \eqref{prob,Laplace} to \eqref{coop}, we arrive at \eqref{mean,fina}, which completes the proof.

\vspace{-4mm}
\section{Proof of Theorem \ref{theorem:nearest}}\label{proof:nearest}
For any node $\vec{x_i}$, we evaluate $\prob\left(B(\vec{x}_{i,n})=1\right)$ as
\begin{align}\label{near1}
\prob\left(B(\vec{x}_{i,n},\Phi)=1\right) = &\;\int_{|\vec{r_2}|=0}^{R_1}\int_{\psi=0}^{2\pi}\prob\left(B(\vec{r_2},\Phi)=1\right)\frac{g_{n}(r(\vec{x_i}))}{2\pi r(\vec{x_i})}|\vec{r_2}|\,d|\vec{r_2}|\,d\psi,
\end{align}
where $\vec{r_2}$ is a vector from $(0,0)$ to a point within the population circle $S_1$ and $\psi$ is the supplement of the angle between $\vec{r_2}$ and $\vec{x_i}$, $r(\vec{x_i})$ is the distance between $\vec{r_2}$ and $\vec{x_i}$, i.e., $r(\vec{x_i})=\sqrt{|\vec{r_2}|^2+|\vec{x_i}|^2+2|\vec{r_2}||\vec{x_i}|\cos\psi}$, and $g_{n}(r)$ is the probability density function (PDF) of distance $r$ given by $g_{n}(r) = \frac{2}{\Gamma(n)}{(\lambda\pi)}^{n}r^{2n-1}\exp(-\lambda\pi r^{2})$ \cite[eq.~(2.12)]{Haenggi:2012:SGW:2480878}.
Applying \eqref{near1} to \eqref{Prnear}, we rewrite $P(n)$ as
\begin{align}\label{near2}
P(n)={\E}_{\Phi}\Bigg\{\!\sum_{\vec{x_i}\in\Phi(\lambda)}\{\prob\left(B(\vec{x_i},\Phi)=1\right)\!\int_{|\vec{r_2}|=0}^{R_1}\!\int_{\psi=0}^{2\pi}\prob\left(B(\vec{r_2},\Phi)=1\right)\frac{g_{n}(r(\vec{x_i}))}{2\pi r(\vec{x_i})}|\vec{r_2}|\,d|\vec{r_2}|\,d\psi\}\!\Bigg\}.
\end{align}

Using ${\E}_{\Phi}\Big\{\sum_{x\in\Phi} h(x,\Phi)\Big\}= \lambda\int_{\mathbb{R}^2}{{\E}_{\Phi}}(h(x,\Phi))dx$ and \eqref{prob,app1}, we rewrite \eqref{near2} as \eqref{near3}.

\section{Numerical Results for Remark \ref{remark:move}}\label{proof:move}
\begin{figure}[H]
\centering
\includegraphics[width=0.6\textwidth]{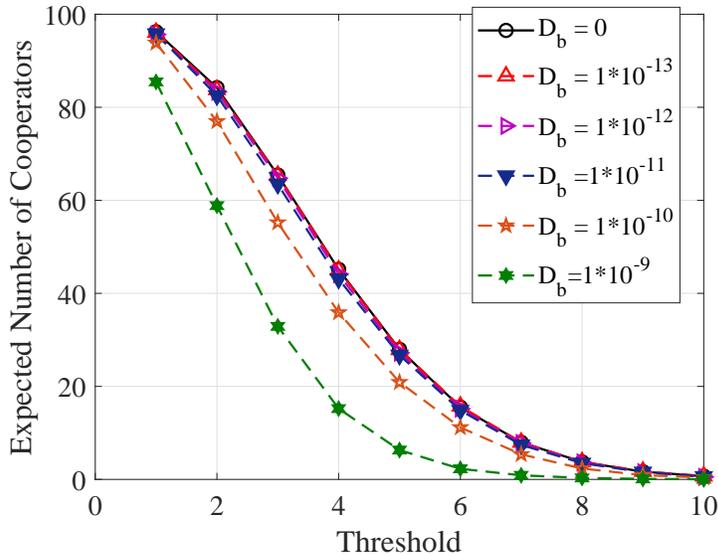}
\caption{Expected number of cooperative bacteria versus detection threshold for different diffusion coefficients of mobile bacteria $D_b$. $R_1 = 50\,{\mu}\m$, expected number of bacteria is 100, and other parameters are the same as those in Section \ref{sec:Numerical}.}
\label{fig:move}
\end{figure}
In Fig. \ref{fig:move}, we plot expected number of cooperative bacteria versus detection threshold for different diffusion coefficients of mobile bacteria $D_b$. We see that the expected number of cooperative bacteria decreases as the diffusion coefficient increases (i.e., bacteria move faster), which meets the Reviewer's expectation (i.e., the cooperation probability is less). This is because when bacteria move faster, the average distance between bacteria is larger and fewer molecules are observed, which leads to a smaller cooperation probability and fewer cooperators. We also see that in comparison to the non-mobile case (i.e., $D_b=0$), the impact of increasing $D_b$ on reducing the number of cooperators is not linear. For example, increasing $D_b$ from 0 to $D_b=10^{-11}$ leads to a negligible drop in the number of cooperators, while increasing $D_b$ from $D_b=10^{-11}$ to $D_b=10^{-9}$ leads to $66\%$ drop of the number of cooperators when the threshold at $\eta=4$. These values may be large for bacteria to diffuse as they are on the order of what we expect for small molecules in water. Though perhaps bacteria can ``diffuse'' this fast via more active means, e.g., via chemotaxis.

\end{appendices}

\vspace{-2mm}


\end{document}